\documentclass[a4paper,11pt]{article}
\pdfoutput=1 
\usepackage{jheppub}
\usepackage[T1]{fontenc} 
\usepackage[utf8]{inputenc}
\usepackage{amsmath}
\usepackage{amssymb}
\usepackage{mathtools}
\usepackage{amsfonts}
\usepackage{bbold}
\usepackage{enumerate}
\usepackage{tikz}
\usetikzlibrary{angles,quotes}
\usepackage{graphicx}
\usepackage{caption}
\usepackage{subcaption}
\usepackage{float}
\usepackage{hyperref}
\usepackage{ulem}
\usepackage{hyperref}
\usepackage{color}
\usepackage{adjustbox}

\begin{document}
\title{Classifying the non-time-local and entangling dynamics of an open qubit system}

\author[]{Sean Prudhoe,}
\emailAdd{stp5134@psu.edu}
\author[]{Sarah Shandera}
\emailAdd{ses47@psu.edu}
\affiliation[a]{Institute for Gravitation and the Cosmos, The Pennsylvania State University, University Park, PA 16802, USA}
\affiliation[b]{Department of Physics, The Pennsylvania State University, University Park, PA, 16802, USA}

\abstract{
 We study families of dynamical maps generated from interactions with varying degrees of symmetry. For a family of time-independent Hamiltonians, we demonstrate the relationship between symmetry, strong-coupling, perfect entanglers, non-Markovian features, and non-time-locality. We show that by perturbing the initial environment state, effective time-local descriptions can be obtained that are non-singular yet capture essential non-unitary features of the reduced dynamics. We then consider a time-dependent Hamiltonian that changes the degree of symmetry by activating a dormant degree of freedom. In this example we find that the one-qubit reduced dynamics changes dramatically. These results can inform the construction of effective theories of open systems when the larger system dynamics is unknown.}

 \arxivnumber{2201.07080}

\maketitle 
\flushbottom

\section {Introduction}

The evolution of quantum systems coupled to unobserved or unobservable degrees of freedom can be much more complex than the evolution of closed systems \cite{OpenQS}. Information may flow back and forth between the observed and unobserved parts of the system, leading to equations of motion that may not be local in time, and that give rise to non-unitary, non-Markovian evolution \cite{Rivas2014}. While formal expressions for the evolution of open systems exist, and exact expressions can be derived in particular cases where the unobserved physics is known, there is not yet a procedure for systematically constructing effective theories of open systems that can encapsulate the full range of possible phenomena. 

A particular class of complex open systems appears in cosmology, where the volume of the universe accessible to a single observer is generically bounded. An open-systems approach in cosmology has been advocated by some researchers for decades \cite{CalzettaHu1989,Koks1997}, and recent theoretical developments have led to a resurgence of interest in these ideas. For example, the possibility of detecting primordial non-Gaussianity in the statistics of the cosmic microwave background fluctuations \cite{Achucarro:2022qrl} motivated to a comprehensive study of the effects of interacting fields during or after inflation. Those interactions can couple Fourier modes of different wavelengths in different ways. The finite extent of the observable universe means that very long wavelength modes are irretrievably in the `environment'. Modes with very short wavelength are practically unobservable as well. Tracing out either, or both, sets of unobservable modes generates an effective description for the observable modes, which form an open system \cite{Shandera:2017qkg}. The appropriate framework for understanding the space of models that generate the data is that of an open effective theory. Moreover, the field or fields relevant for inflation are expected to be accompanied by many other degrees of freedom which may move from passive to active as inflation proceeds. Indeed, it is postulated that if the inflationary field explores too large of a range, a large number of fields will become relevant, limiting the validity of the original model as an effective description \cite{Ooguri:2006in, Garg:2018reu, Ooguri:2018wrx}. Using an open system effective theory for the inflaton, rather than a traditional low-energy effective theory, one can treat these light degrees of freedom as a time-dependent environment. These open systems are complex, since many results that are known require non-perturbative techniques \cite{Burgess:2009bs} and the associated open systems are frequently non-Markovian \cite{Agon2018}. However, one has symmetry constraints to guide the effective theory \cite{Weinberg:2008hq}. A similar story of complexity restricted by symmetry applies to another well-studied system where open effective theories are applicable, black holes \cite{Banks:1983by, Unruh:1995gn, Burgess:2014eoa,Burgess:2015ajz,Kaplanek:2019vzj, Kaplanek:2020iay, Burgess:2021luo}. Cosmologists, then, would like to understand how to construct effective theories of open systems that are non-perturbative and non-Markovian, with time-dependent environments, but constrained by symmetries.

To address these questions requires a non-perturbative understanding of the evolution of the reduced system. For finite-dimensional quantum systems this can be addressed using the dynamical map, the operator that governs the dynamics of the reduced degrees of freedom. Of course, determining the dynamical map is as difficult as solving the Schr\"{o}dinger equation governing the system/environment dynamics. Thus for this work we consider simple open systems, constructing dynamical maps for a single system qubit. In this setting the dynamical map can be determined as an analytic function of the Hamiltonian parameters, for a generic initial environment state. 

Due to their relative simplicity, much is understood about the general structure of qubit dynamical maps \cite{nielsen00,Ruskai2001,Braun_2014,Rodriguez06}. Such maps appear frequently in quantum computational settings, where they model noisy interactions of qubits with the environment.  Optimal dynamical maps have been found for performing communication tasks that leverage qubit entanglement to transmit quantum information securely  \cite{Niu_1999,King_2001,Berry_2005,https://doi.org/10.48550/arxiv.2006.16366,Ghosal_2021}. So, while interacting qubits are far simpler systems than those found in cosmology, it is an appropriate starting point to connect the questions of interest to cosmologists to the recent advances in the understanding of open systems that have been driven by laboratory and quantum computational considerations.

We begin by investigating a solvable example of coupled qubits to explore how non-Markovianity and the non-time-local aspects of the master equation for one qubit depend on the initial state of the unobserved qubit and on the symmetries and coupling constants of the full Hamiltonian. The first Hamiltonian we consider has (1) a conserved quantity, and (2) a block-diagonal structure of two equal-size pieces, allowing additional symmetry structures at special points in parameter space. Since non-time-local master equations can be particularly difficult to work with, we examine how much of the parameter space requires a time non-local equation, and the perturbations or approximations in both the Hamiltonian or the state of the unobserved qubit that will generate time-local equations of motion. 

We then extend the study of the two qubit model, using a subset of the Hamiltonian family and initial environment states that allow for the appearance of additional degree of symmetry. We introduce a third qubit that remains a spectator for a time and is later switched on using an interaction that explicitly breaks the extra symmetry appearing in the two qubit interaction. The study  focuses on the radical change in the invertibility of the dynamical map obtained by tracing out the two qubits that act as the environment, indicating that a time-local master equation cannot be used after symmetry breaking. 

In the rest of the introduction we briefly review the formalism for open system dynamics and master equations in the context of our goals and model. Then, in Section \ref{sec:ExactDynam}, we introduce the example systems. In Section \ref{sec:reducedDynamics} we derive the reduced dynamics for a single-qubit systems, tracing out the environment qubit(s). Section \ref{sec:features} derives the conditions for non-time-local dynamics via the non-invertibility of the dynamical map and discusses several features we use to classify the dynamical map including the non-Markovianity, divisibility, and symmetries of the Hamiltonian. We determine how these depend on Hamiltonian parameters, in particular whether it is strongly coupled or not, and what role the initial environment state plays in non-time-locality. In Section \ref{sec:mastereqn} we use perturbed initial environment states to construct approximate time-local master equations. And we conclude in Section \ref{sec:conclude}. 

\subsection{Open system evolution}
\label{subsec:master_eqns}
For some open systems, the master equation governing the evolution of the density matrix for the observed system, $\rho_{\rm S}(t)$, is \cite{Lindblad,Vittorio:1976,Shibata:1977,Chaturvedi:1979}
\begin{align}
    \partial_{t} \rho_{\rm S}(t) 
    &= -i[H_{\rm free}(t)+H_{\rm open}(t),\rho_{\rm S}(t)]+\sum_{k}\gamma_{k}(t)\left(L_{k}\rho_{\rm S}(t)L^{\dag}_{k}-\frac{1}{2}\{L^{\dag}_{k}L_{k},\rho_{\rm S}(t)\}\right)\,.
    \label{eq:Lindblad}
\end{align}
Here $H_{\rm free}(t)+H_{\rm open}(t)=H_{\rm eff}(t)$ is the effective Hamiltonian of the system, containing both the original system Hamiltonian $H_{\rm free}(t)$ and a piece, $H_{\rm open}(t)$, generated by the coupling to an environment. The $L_k$ are operators acting on the system, and the $\gamma_{k}(t)$ are functions describing the flow of information between the system and environment. The $\gamma_{k}(t)$, which control the subset of all possible operators $L_k$ that appear with non-zero coefficients, depend on the system-environment coupling and the state of the environment. $H_{\rm eff}(t)$ and $\gamma_{k}(t)$ are given by environment correlation functions calculated using the initial environment state $\rho_{\rm E}(0)$.

If the environment and full Hamiltonian are unknown, one might begin constructing an effective theory for the system by writing all possible $L_k$ and a generic $H_{\rm eff}$ from the complete set of operators that act on the system. Then, the work in the effective theory comes in specifying some structure for the dissipation functions $\gamma_k(t)$, determining any approximations that may allow some possible terms in $H_{\rm eff}$ to be discarded, and in determining consistency between effects captured in $H_{\rm open}(t)$ and in the non-unitary part of Eq.(\ref{eq:Lindblad}). Some broad guidelines for this process are known: the simplest choice would be all $\gamma_k\geq 0$ and constant, restricting the system to non-unitary however time-independent, Markovian evolution. Time-dependent Markovian dynamics would be described by $\gamma_k(t)\geq0$ at all times. Finally a restricted set of non-Markovian dynamics would be captured by considering generic functions $\gamma(t)$. 

However, the most general case allows the master equation to be non-local in time. Then, in addition to the time-local part one adds an integral term. This is the Nakajima-Zwanzig equation \cite{1958PThPh..20..948N,1960JChPh..33.1338Z}, 
\begin{equation}
    \partial_{t}\rho_{\rm S}(t)= K_{\rm TL}(t)\rho_{\rm S}(t)+\int K_{\rm NZ}(t,\tau)\rho_{\rm S}(\tau)d\tau\,,
    \label{eq:tnonlocal}
\end{equation}
where the time-local piece, $K_{\rm TL}(t)$, generates the same action on $\rho_{\rm S}(t)$ as given in Eq.(\ref{eq:Lindblad}). The integral is over the history of the evolution $[t_{0},t]$ where $t_{0}$ is a time where the system and environment are uncorrelated and $t$ is the time where one is interested in calculating ensemble averages.  In constructing a parameterized effective theory for the open system dynamics, one would like to know how to systematically address whether a non-local equation is necessary. In addition, which qualitative aspects of information flow can be captured in either the $\gamma(t)$ or the time non-local kernel $K_{\rm NZ}(t,\tau)$, and how they should be implemented? 

To address these questions in a simple case, we explore the relationship between the full Hamiltonian for system and environment, together with the initial state of the environment, to several features of the reduced dynamics. We consider measures of non-Markovianity and the conditions under which time non-local evolution is required. We do this by first computing the exact reduced dynamics via the dynamical map. This is a non-unitary generalization of the time evolution operator, a completely positive and trace preserving (CPTP) map from the initial density matrix to the density matrix at a later time $t$, 
 \begin{equation}
    \rho_{S}(t) = \Lambda(t,0) \circ \rho_{S}(0)\,.
    \label{eq:dynmapIntro}
\end{equation}
For qubit dynamical maps the complications arising from non-Markovianity are less severe. Non-Markovian qubit dynamical maps can be tractably studied, more so than their master equation counterparts. For example the divisibility (how the time evolution can be broken into steps) and how it relates to non-Markovianity has been exhaustively studied in the case of qubit dynamical maps \cite{2019A,2019B,2019C,Chru_ci_ski_2014}. For the model we consider it is therefore possible to make detailed statements about the relation between properties of the dynamical map and its non-Markovianity. 

Dynamical maps do not have to be invertible; maps with ${\rm Det}\Lambda(\tau_{i})=0$ for some times $\tau_{i}$ require either a time-local description that diverges at each $\tau_{i}$ or a non-time-local integral kernel, as written in Eq.(\ref{eq:tnonlocal}) \cite{Chru_ci_ski_2010}. For a simple system, we will use the non-invertibility of the dynamical map to derive the conditions on the full Hamiltonian and the environment that make a non-time-local master equation necessary. We find that, independent of the Hamiltonian, there are always a set of initial environment states which support time-local reduced dynamics. This allows a time-local, approximate, master equation to be constructed by shifting the initial state of the environment. For an initial environment state, $\rho_{\rm E}(t_0)$, and dynamics that requires a non-time-local piece, there exist environment states nearby in trace distance norm, $\{\rho^{'}_{\rm E}(t_0)\}$, which can be used to define a master equation of the form
\begin{equation}
\begin{split}
\partial_{t}\rho_{S}(t;\rho_{\rm E}(t_{0}))&=K_{\rm TL}(t;\rho^{'}_{\rm E}(t_{0}))\rho_{S}(t;\rho_{\rm E}(t_{0}))\\&+\int K_{\rm NTL}(t,\tau;\delta\rho_{\rm E})\rho_{\rm S}(\tau;\rho_{\rm E}(t_{0}))d\tau\,.
\end{split}
\end{equation}
Here the non-time-local component is linear in $\delta\rho_{\rm E}=\rho_{\rm E}(t_{0})-\rho^{'}_{\rm E}(t_{0})$, and the integral is over the evolution history $[t_{0},t]$.
  
\section{The system-environment Hamiltonians and the unitary dynamics}
\label{sec:ExactDynam}
This section introduces both the family of two-qubit Hamiltonians considered, as well as a more restrictive family of three-qubit interactions. The two-qubit model has an associated parity symmetry, which splits the Hamiltonian into two equal-sized blocks. Such models are interesting as they have an intermediate level of symmetry: more than the class with no non-trivial symmetries, however not as much as Hamiltonians that preserve the total angular momentum of the two qubits. Physically, dynamics of the type we use here describe a pair of non-interacting qubits, most clearly seen through a change of meronomic frame \cite{Hulse:2019}. We also characterize features of the non-Markovianity of the single-qubit evolution, according to parameter choices in the full Hamiltonian.

\subsection{Two-qubit family}
The two-qubit Hamiltonian that we study is 
 \begin{equation}
 \begin{split}
      H=&H_{\rm free}+H_{\rm int}\\=&\omega_{\rm S}(Z_{\rm S}\otimes\mathbb{1}_{\rm E})+\omega_{\rm E}(\mathbb{1}_{\rm S}\otimes Z_{\rm E}) 
       +\kappa_{\rm SE}(Y_{\rm S}\otimes X_{\rm E})+\kappa_{\rm ES}(X_{\rm S}\otimes Y_{\rm E}) \,.
\end{split}
\label{eq:H}
\end{equation}
The free parameters $\omega_{\rm S}$ and $\omega_{\rm E}$ provide the time scales associated to the free dynamics of each individual qubit (with $\hbar=1$), and the parameters $\kappa_{SE}$ and $\kappa_{ES}$ are coupling strengths. $H$ has a symmetry, $[H, Z_{\rm S}\otimes Z_{\rm E}]=0$, so the eigenstates have definite parity associated to $P_{zz}=Z_{\rm S} \otimes Z_{\rm E}$. This $\mathbb{Z}_{2}$ symmetry allows $H$ to be split into even and odd blocks, where the even block is spanned by states with correlated spins (e.g. $|\uparrow\uparrow\rangle$) and the odd block is spanned by states with anti-correlated spins (e.g. $|\uparrow \downarrow \rangle$).

The block diagonalization is achieved by splitting $H$ into symmetric and anti-symmetric parts under exchange of system and environment operators. Defining new parameters,
\begin{equation}
\begin{split}
    &2\Delta_{\pm}=\omega_{\rm S}\pm \omega_{\rm E} \\ 
    &2\kappa_{\pm}=\kappa_{\rm SE}\pm \kappa_{\rm ES}\,,
\end{split}
\end{equation} 
the Hamiltonian can be written as
\begin{equation}
\begin{split}
    H= &\Delta_{+}(Z_{\rm S}\otimes\mathbb{1}_{\rm E}+\mathbb{1}_{\rm S}\otimes Z_{\rm E})+\kappa_{+}\left[Y_{\rm S}\otimes X_{\rm E}+X_{S}\otimes Y_{\rm E}\right] \\
    +&\Delta_{-}(Z_{\rm S}\otimes\mathbb{1}_{\rm E}-\mathbb{1}_{\rm S}\otimes Z_{\rm E})+\kappa_{-}\left[Y_{\rm S}\otimes X_{\rm E}-X_{\rm S}\otimes Y_{\rm E}\right]\\
    \equiv&H_{+}+H_{-} \, ,
\end{split}
\label{eq:HSEblocks}
\end{equation} 
 where the $\pm$ labels correspond to the $\mathbb{Z}_{2}$ (parity) eigenvalues of each block. The block diagonalization of $H$ introduces a subspace decomposition $\mathcal{H}=\mathcal{Q}_{+}\oplus \mathcal{Q}_{\rm -}$, where the spaces $\mathcal{Q}_{\pm}$ are spanned by the eigenstates of $H_{\pm}$. As we show below, there is also a tensor product decomposition for which the two subsystems decouple. 

Using $|0\rangle$ and $|1\rangle$ to label the eigenstates of $Z_{\rm S}$ and $Z_{\rm E}$, the stationary states of $H$ are
\begin{equation}
    \begin{split}
        &|0_{+} \rangle =\cos{\frac{\phi_{+}}{2}}|0_{\rm S},0_{\rm E}\rangle +i\sin{\frac{\phi_{+}}{2}}|1_{\rm S},1_{\rm E}\rangle\\
        &|1_{+}\rangle =\sin{\frac{\phi_{+}}{2}}|0_{\rm S},0_{\rm E}\rangle -i\cos{\frac{\phi_{+}}{2}}|1_{\rm S},1_{\rm E}\rangle \\ 
        &|0_{-}\rangle=\cos{\frac{\phi_{-}}{2}}|0_{\rm S},1_{\rm E}\rangle +i\sin{\frac{\phi_{-}}{2}}|1_{\rm S},0_{\rm E}\rangle\\
        &|1_{-} \rangle =\sin{\frac{\phi_{-}}{2}}|0_{\rm S},1_{\rm E}\rangle -i\cos{\frac{\phi_{-}}{2}}|1_{\rm S},0_{\rm E}\rangle \,,
     \end{split}
  \label{eq:eigen1}
\end{equation}
with eigenvalues
\begin{equation}
    \pm\omega_{\pm}=\pm 2\sqrt{\Delta_{\pm}^{2}+\kappa_{\pm}^{2}}\,.
\end{equation}
In Equation (\ref{eq:eigen1}), the angles
 \begin{equation}
        \phi_{\pm}= \arctan{\frac{\kappa_{\pm}}{\Delta_{\pm}}}
\label{eq:phi}
\end{equation}
indicate the relative size of the interaction and free Hamiltonian parameters, and if either provide a dominate contribution to the energy eigenvalues $\omega_{\pm}$. As long as at least one of the blocks is interacting i.e. $\kappa_{\pm} \neq 0$, the stationary states in the subsystem decomposition $\mathcal{H}=\mathcal{Q}_{\rm S}\otimes \mathcal{Q}_{\rm E}$ are entangled, and maximally entangled as $\frac{\kappa_{\pm}}{\Delta_{\pm}} \rightarrow \pm \infty$.

 We will see below that the reduced dynamics may acquire a non-time-local component if $\phi_{+}+\phi_{-} \geq \frac{\pi}{2}$. Since this condition requires that one or both of $\kappa_{\pm}\geq \Delta_{\pm}$, the part of parameter space where non-time-local master equations can be required coincides with strong coupling, although not all strongly coupled Hamiltonians will have non-time-local dynamics.

So far, we have defined subsystems assuming a laboratory-based notion of locality for operations on qubits, established by the system/environment labels. However, the block-diagonal structure of the Hamiltonian suggests that we also consider a re-organization of the Hilbert space into degrees of freedom that decouple. That is, we can define qubits $A$ and $B$, with orthonormal basis states $\{1_{\rm A}, 0_{\rm A}\}$ and  $\{1_{\rm B}, 0_{\rm B}\}$, so that $\mathcal{H}=\mathcal{Q}_{\rm A}\otimes \mathcal{Q}_{\rm B}$. This is a change of meronomic frame \cite{Hulse:2019}. Explicitly, a (non-unique) mapping between these bases is given by
\begin{equation}
\begin{split}
    |0_{+}\rangle=&|0_{\rm A},0_{\rm B}\rangle\\
    |1_{+} \rangle=&|1_{\rm A},1_{\rm B}\rangle\\
    |0_{-}\rangle=&|0_{\rm A},1_{\rm B}\rangle\\
    |1_{-} \rangle=&|1_{\rm A},0_{\rm B}\rangle \, .
\end{split}
\label{eq:newFrame}
\end{equation}

 One finds $H=\omega_+\left(|0_{+} \rangle\langle 0_{+} |-|1_{+}\rangle\langle 1_{+}|\right)+\,\omega_-\left(|0_{-} \rangle\langle 0_{-} |-|1_{-}\rangle\langle 1_{-}|\right)$, it is straightforward to change to the frame given in Eq.(\ref{eq:newFrame}), where 
\begin{equation}
H_{AB} =\omega_{\rm A}Z_{\rm A}\otimes\mathbb{1}_{\rm B}+\omega_{\rm B}\mathbb{1}_{\rm A}\otimes Z_{\rm B}\,.
   \label{eq:HAB}
   \end{equation}
Here $\omega_{\rm A}=\frac{1}{2}(\omega_{+}+\omega_{-})$, $\omega_{\rm B}=\frac{1}{2}(\omega_{+}-\omega_{-})$, and $Z_{\rm A}=|0_{\rm A}\rangle\langle 0_{\rm A}|-|1_{\rm A}\rangle\langle 1_{\rm A}|$, etc. Comparing to Eq.(\ref{eq:HSEblocks}), the two terms in Eq.(\ref{eq:HAB}) are just the A/B frame expressions for $H_+$ and $H_-$. An advantage of this frame is that it is easy to characterize the regions of parameter space with extra symmetry. A particularly useful region of parameter space is 
\begin{equation}
   \;\;\;\;\;\;\;\; \omega_{+}=\omega_{-} \Leftrightarrow\omega_{\rm B}= 0\,\,\, {(\rm \bf the \,\, degenerate\,\,family)}\,
       \label{eq:degenerateDef}
\end{equation}
which is symmetric under local rotations on the B qubit subsystem. Since symmetries are clearest in the A/B frame, we will continue to use it for that purpose in Table \ref{table} below, where we classify the reduced system dynamics possible with this Hamiltonian, Eq.(\ref{eq:H}).

Returning to the system/environment frame, we characterize the entangling properties of the time evolution. Using the computational basis for system and environment qubits, the time evolution operator, $U(t)$, is 
\begin{equation}
    \begin{split}
        U(t) =e^{-iHt}= \begin{bmatrix}  \alpha_{+}(t) & 0 &0& -\beta_{+}(t) \\ 
        0& \alpha_{-}(t) &-\beta_{-}(t)&0 \\ 0&\beta_{-}(t)&\bar{\alpha}_{-}(t)&0 \\ \beta_{+}(t) &0&0& \bar{\alpha}_{+}(t) \end{bmatrix} \\,
    \end{split}
\end{equation}
where
\begin{equation}
    \begin{split}
    &\alpha_{\pm}(t) = \cos\omega_{\pm}t-i\cos\phi_{\pm}\sin\omega_{\pm}t\,, \\ 
    &\beta_{\pm}(t) = \sin\phi_{\pm}\sin\omega_{\pm}t \,,
    \end{split}
    \label{eq:ABs}
\end{equation}
and $\bar{\alpha}_{\pm}(t)$ is the complex conjugate of $\alpha_{\pm}(t)$. The pair of functions from each doublet satisfy $|\alpha_{\pm}(t)|^{2}+|\beta_{\pm}(t)|^{2}=1$. 

The functions $\beta_{\pm}(t)$ are generated by the interaction between system and environment, so they determine both the entanglement and, as we see below, the invertibility of the reduced dynamics for the system.

\subsection{Three-qubit model of dynamical symmetry breaking}
We further study a model where the reduced dynamics is phase covariant.   
 Reduced dynamics is phase covariant or time translation symmetric if the following constraints hold \cite{https://doi.org/10.48550/arxiv.2106.05295}
\begin{enumerate}[I]
\item $[H_{\rm free},H_{\rm int}]=0$
\item $[H_{\rm free},\rho_{\rm E}(0)]=0$ .
\end{enumerate}
These constraints have thermodynamic implications. The first constraint is a strict energy conservation condition i.e. energy is not built up between the system and environment boundary. The second assumption is that the initial environment state is a Gibbs' state defined with respect to the free environment Hamiltonian. If these assumptions hold, then the set of time evolution operator generated by the free system Hamiltonian commutes with the action of the dynamical map. Hence the designation as a time translation symmetry.

To incorporate phase-covariance breaking into our model we consider the Hamiltonian 
\begin{equation}
   \begin{split}
 \tilde{H}(t)=& H_{\rm free}+H_{\rm int}(t) \\ &\omega(Z_{\rm S}\otimes\mathbb{1}_{\rm E}\otimes\mathbb{1}_{\rm E '}+\mathbb{1}_{\rm S}\otimes Z_{\rm E}\otimes \mathbb{1}_{\rm E'})+\omega\left[ X_{S}\otimes Y_{\rm E}\otimes\mathbb{1}_{\rm E'}-Y_{\rm S}\otimes X_{\rm E}\otimes\mathbb{1}_{\rm E'}\right] \\ &+\gamma\Theta(t-\tau)\left[X_{\rm S}\otimes\mathbb{1}_{\rm E}\otimes X_{\rm E '}\right] \,,
 \end{split}
\end{equation}
where $\Theta(t)$ is the Heaviside theta function. The choice of the time-independent portion of the Hamiltonian satisfies strict energy conservation. The second assumption is satisfied if the initial state of the E qubit is restricted to the form $\rho_{\rm E}(0)=\frac{1}{2}(\mathbb{1}_{\rm E}+z_{\rm E}(0)Z_{\rm E})$. For what follows $\tilde{H}_{t<\tau}$ is the Hamiltonian for $t<\tau$ and $\tilde{H}_{t\geq \tau}$ is the Hamiltonian for $t\geq \tau$, as we can note the Hamiltonian is a piece-wise define function.  

$\tilde{H}_{t\geq \tau}$ is chosen to break the phase covariance of the reduced dynamics, but in the process also breaks the parity symmetry. But other global symmetries appear for $t \geq \tau$, which we can exploit to determine the stationary states when the interaction changes. Using the Pauli anti-commutation relations, it is not difficult to see $[\tilde{H}_{t\geq \tau},Z_{\rm S}\otimes Z_{\rm E}\otimes Z_{\rm E'}]=0$. But from the block structure in the eigendecomposition of the Hamiltonian, one can note the presence of another symmetry. We find that the observable, 
\begin{equation}
\begin{split}
\mathcal{O}_{2}=Y_{\rm S}\otimes\mathbb{1}_{\rm E}\otimes Y_{\rm E'}-Z_{\rm S}\otimes X_{\rm E}\otimes Y_{\rm E'} \,
\end{split}
\end{equation}
commutes with $\tilde{H}_{t\geq\tau}$. Therefore $\tilde{H}_{t\geq \tau}$ splits into four $2\times2$ blocks with definite values of $\mathcal{O}_{1}$ and $\mathcal{O}_{2}$

Thus a familiar path may be taken to determine the time evolution generated by $\tilde{H}_{t\geq\tau}$, which we denote by $\tilde{U}_{t\geq\tau}(t)$. We begin with the eigenvalues which are found to be 
\begin{equation}
    \begin{split}
\pm\Omega_{\pm}=\pm\sqrt{\gamma^{2}\pm2\sqrt{2}\gamma\omega+4\omega^{2}}
    \end{split}
\end{equation}
and are doubly degenerate. We shall start the computation in the eigenbasis of $\mathcal{O}_{2}$ (called the polarized basis), and then go back to the computational basis. To this end define the angles, 
\begin{equation}
\tan\psi_{\pm}=1\pm 2\sqrt{2}\frac{\omega}{\gamma}
\end{equation}
from which we obtain the time dependent functions (note the similarity to the previous section), 
\begin{equation}
\begin{split}
    &\tilde{\alpha}_{\pm}(t)=\cos\Omega_{\pm}t-i\cos\psi_{\pm}\sin\Omega_{\pm}t \\
    &\tilde{\beta}_{\pm}(t)=\sin\psi_{\pm}\sin\Omega_{\pm}t
\end{split}
\end{equation}
With these functions one can construct the unitary operator  $\tilde{U}_{t\geq \tau}(t)$ in the polarized basis. In the computational basis, the following combinations of functions will be most useful 
\begin{equation}
\label{eq:abgh}
    \begin{split}
        &a_{\pm}(t)=\frac{1}{2}\Re{(\tilde{\alpha}_{+}\pm\tilde{\alpha}_{-})} \\
        &b_{\pm}(t)=\frac{1}{2}\Im{(\tilde{\alpha}_{+}\pm\tilde{\alpha}_{-})}\\ 
        &g_{\pm}(t)=\frac{1}{2}(\tilde{\beta}_{+}\pm\tilde{\beta}_{-})  \\
        &h_{\pm}(t)=\frac{1}{2}(\tilde{\alpha}_{+}\pm\bar{\tilde{\alpha}}_{-})
    \end{split}\,.
\end{equation}
We have all we need to compute the total time evolution operator, which is given by the following time ordered exponential 
\begin{equation}
    \tilde{U}(t,0)=\mathcal{T}\exp{\left[-i\int_{0}^{t}\tilde{H}(\tau)d\tau\right]} \,.
\end{equation}
 Breaking the time interval into $N$ pieces, and using the Baker-Campbell-Hausdorff relation, one can show that in the limit $N$ goes to infinity the time ordered exponential approaches 
\begin{equation}
\begin{split}
     \tilde{U}(t,0)=&\left[\Theta(t)-\Theta(t-\tau)\right]\tilde{U}_{t<\tau}(t)+\Theta(t-\tau)\tilde{U}_{t\geq \tau}(t-\tau)\tilde{U}_{t <\tau}(\tau) 
\end{split}
\end{equation}
where we have just seen how to compute $\tilde{U}_{t\geq \tau}(t)$, and we can use the results of the previous section to determine $\tilde{U}_{t<\tau}(t)$. 

\section{The reduced dynamics }
\label{sec:reducedDynamics}
The reduced dynamics of the system is obtained from the full dynamics by tracing out the environment. Restricting to factorized initial states ensures that the reduced dynamics is completely positive \cite{Lindblad}. In that case, the density matrix of the system qubit alone, at time $t$, is 
\begin{equation}
 \rho_{\rm S}(t) = tr_{\rm E} [U(t)  \rho_{\rm S}(0) \otimes \rho_{\rm E}(0)  U^{\dag}(t)]\,.
 \label{eq:TE}
\end{equation}
 
The state of the system qubit at any time can be found using the dynamical map, represented by a matrix acting on the vectorization of the reduced density matrix \cite{Rivas2014}. For a qubit density matrix this is the Bloch representation
\begin{equation}
    \rho_{\rm S}(0)= \frac{1}{2}(\mathbb{1}_{\rm S}+ \vec{r}_{\rm S}(0)\cdot \vec{\sigma}_{\rm S}),
\end{equation}
where $\vec{r}_{\rm S}(0)$ is a real, 3-dimensional vector with Euclidean norm $||\vec{r}_{\rm S}(0)||\leq 1$.
. 
The evolved density matrix can then be written as
\begin{equation}
    \rho_{\rm S}(t) = \Lambda(t,0) \circ \rho_{\rm S}(0).
    \label{eq:dynmap}
\end{equation}
where $\Lambda(t,0)$ is a $4\times4$ matrix. Note that as the Hamiltonian is time independent, the dynamical map depends on the initial and final times only through the difference $t-t_{0}$, so for what follows we can simply write $\Lambda(t)$ and suppress the initial time dependence. It is helpful to further define a $3\times3$ matrix with components $T^{ij}=\Lambda^{ij}$ and a 3-vector $d^{i}=\Lambda^{0i}$ \cite{Ruskai2001}, so that the action of the dynamical map can be written 
 \begin{equation}
 \begin{split}
    \rho_{\rm S}(t) =& \frac{1}{2}(\mathbb{1}_{\rm S}+\vec{r}_{\rm S}(t)\cdot \vec{\sigma}_{\rm S})  \\ =&\frac{1}{2}\left(\mathbb{1}_{\rm S}+\left(\vec{\vec{T}}(t)\circ\vec{r}_{\rm S}(0)+\vec{d}(t)\right)\cdot \vec{\sigma}_{\rm S}\right)\,.
\end{split}
\label{eq:Action}
\end{equation}
This presentation has the advantage that the trace fixing requirement is immediate. For $\Lambda$ to be physical, $\vec{\vec{T}}$ and $\vec{d}$ must be real so that $\rho_{\rm S}(t)$ is Hermitian. Unlike unitary maps, dynamical maps need not be divisible. For example, one is generally unable to split the time evolution as $\Lambda(t+\tau)=\Lambda(\tau)\Lambda(t)$. The reduced dynamics further differs from unitary dynamics by exhibiting hallmark features of open systems, including purity change and decoherence.

 As with ordinary operators, we can define components of the  dynamical map with respect to an operator basis on $\mathcal{Q}_{\rm S}$,
\begin{equation}
    \Lambda^{ab}(t)= \frac{1}{2}tr_{\rm S}[\sigma^{a}_{\rm S} \Lambda(t)\circ \sigma^{b}_{\rm S}] = \frac{1}{2}tr[(\sigma^{a}_{\rm S}\otimes\mathbb{1}_{\rm E}) U(t)(\sigma^{b}_{\rm S}\otimes\rho_{\rm E}(0))U^{\dag}(t)]
\end{equation}
where $a$ and $b$ can take the values $0,x,y,z$ where $\sigma^{0}_{S} = \mathbb{1}_{\rm S}$. 


\subsection{Two-qubit family}

Carrying out this calculation, we find the non-zero dynamical map components, in terms of the environment qubit's initial state, $\vec{r}_E$(0), and the functions $\alpha_{\pm}(t)$ and $\beta_{\pm}(t)$ appearing in the time evolution operator, Eq.(\ref{eq:ABs}), to be

\begin{equation}
\begin{split}
\Lambda^{z0}(t)=& \frac{1}{2}\left[|\alpha_{+}(t)|^{2}-|\beta_{+}(t)|^{2}-|\alpha_{-}(t)|^{2}+|\beta_{-}(t)|^{2}\right]z_{E}(0) \\ 
    \Lambda^{xx}(t) =& \Re[\alpha_{+}(t)\alpha_{-}(t)-\beta_{+}(t)\beta_{-}(t)]  \\ 
    \Lambda^{xy}(t) =& \Im[\alpha_{+}(t)\alpha_{-}(t)+\beta_{+}(t)\beta_{-}(t)]  \\
    \Lambda^{xz}(t) =& \Re[\alpha_{+}(t)\beta_{-}(t)+\alpha_{-}(t)\beta_{+}(t)]x_{E}(0)-\Im[\alpha_{+}(t)\beta_{-}(t)-\alpha_{-}(t)\beta_{+}(t)]y_{E}(0) \\ 
    \Lambda^{yx}(t) =&-\Im[\alpha_{+}(t)\alpha_{-}(t)-\beta_{+}(t)\beta_{-}(t)]  \\ 
    \Lambda^{yy}(t) =&\Re[\alpha_{+}(t)\alpha_{-}(t)+\beta_{+}(t)\beta_{-}(t)]  \\
    \Lambda^{yz}(t)=&-\Im[\alpha_{+}(t)\beta_{-}(t)+\alpha_{-}(t)\beta_{+}(t)]x_{E}(0)-\Re[\alpha_{+}(t)\beta_{-}(t)-\alpha_{-}(t)\beta_{+ }(t)]y_{E}(0)\\
    \Lambda^{zx}(t)=&-\Re[\alpha_{+}(t)\beta_{+}(t)+\alpha_{-}(t)\beta_{-}(t)]x_{E}(0)+\Im[\alpha_{+}(t)\beta_{+}(t)-\alpha_{-}(t)\beta_{-}(t)]y_{E}(0) \\ 
    \Lambda^{zy}(t)=& -\Im[\alpha_{+}(t)\beta_{+}(t)+\alpha_{-}(t)\beta_{-}(t)]x_{E}(0)-\Re[\alpha_{+}(t)\beta_{+}(t)-\alpha_{-}(t)\beta_{-}(t)]y_{E}(0)\\ 
    \Lambda^{zz}(t)=& \frac{1}{2}\left[|\alpha_{+}(t)|^{2}-|\beta_{+}(t)|^{2}+|\alpha_{-}(t)|^{2}-|\beta_{-}(t)|^{2}\right] \,.
\end{split} 
\label{eq:Lambdacomponents}
\end{equation}
Note the difference in how the components of the initial environment state enter, with $z_{\rm E}(0)$ appearing separately from $x_{\rm E}(0)$ and $y_{\rm E}(0)$. It is instructive to see how the symmetry of the full Hamiltonian simplifies the dynamical map dependence on the initial state of the environment qubit, $\vec{r}_{\rm E}$(0). To that end, define the partial components of the dynamical map by
\begin{equation}
\begin{split}
    \Lambda^{ab}&=\frac{1}{2}\Lambda^{abc}r^{c}_{\rm E}(0) \\
    \Lambda^{abc}&=tr[(\sigma^{a}_{\rm S}\otimes \mathbb{1}_{\rm E})U(t)(\sigma^{b}_{\rm S}\otimes \sigma^{c}_{\rm E})U^{\dag}(t)]\,.
\end{split}
\end{equation} 
The parity symmetry $P_{zz}=Z_{\rm S}\otimes Z_{\rm E}$ essentially halves the number of non-zero $\Lambda^{abc}$, and separates the $xy$ and $z$ components of initial environment state in the dynamical map.

Then, since $P_{zz}$ commutes with $U(t)$ and satisfies $P^{2}_{zz}=\mathbb{1}$, we see that
\begin{equation}
 \begin{split}
    \Lambda^{abc}&=tr[P^{2}_{zz}(\sigma^{a}_{\rm S}\otimes \mathbb{1}_{\rm E})U(t)(\sigma^{b}_{\rm S}\otimes \sigma^{c}_{\rm E})U^{\dag}(t)] \\
                 &=(-1)^{\pi_{a}+\pi_{b}+\pi_{c}}tr[P_{zz}(\sigma^{a}_{\rm S}\otimes \mathbb{1}_{\rm E})U(t)(\sigma^{b}_{\rm S}\otimes \sigma^{c}_{\rm E})U^{\dag}(t)P_{zz}] \\ 
                 &=(-1)^{\pi_{a}+\pi_{b}+\pi_{c}}\Lambda^{abc}
\end{split}\,,
\label{eq:parityComponents}
\end{equation}
where the $\pi_{d}$ are defined such that, 
\begin{equation}
    P_{zz}(\sigma^{e}_{\rm S}\otimes \sigma^{f}_{\rm E})=(-1)^{\pi_{e}+\pi_{f}}(\sigma^{e}_{\rm S}\otimes \sigma^{f}_{\rm E})P_{zz} \,.
\end{equation}
Considering the cases where either e=0 or f=0, allows one to speak of the parity of local system and environment operators. For example the parity of the operator $\mathbb{1}_{\rm S}\otimes\sigma^{f}_{\rm E}$ is determined by $\pi_{f}$. The condition $\pi_{a}+\pi_{b}+\pi_{c}=1$ (mod 2) if satisfied implies that $\Lambda^{abc}$ vanishes. This is equivalent to the following: given $\Lambda^{ab}$ if the parity of $(\pi_{a}+\pi_{b})$ is even (odd) then the only contributing partial components have $\pi_{c}$ even (odd).  $\Lambda^{zzz}$ also vanishes as a consequence of the parity symmetry. However this is obviously not a consequence of the previous argument. Instead it follows using the $P_{zz}$ symmetry and that $tr_{\rm S}(Z_{\rm S})=0$. 

Several additional partial components vanish, although not enforced by Eq.(\ref{eq:parityComponents}). For example,
\begin{equation}
    \begin{split} 
       &\Lambda^{xtx}=\Lambda^{xty}= 0 \\ 
       &\Lambda^{ytx}= \Lambda^{yty}=0 
    \end{split} 
\end{equation}
which imposes that the shift $\vec{d}$ must be parallel to the z axis. Finally, 
\begin{equation}
    \Lambda^{xxz}=\Lambda^{xyz}=\Lambda^{yxz}=\Lambda^{yyz}=0 \,.
\end{equation}

Important to mention is that the shift of the dynamical map is restricted to having only a non-zero z-component.  Thus the dynamical map is unital if $z_{rm E}(0)=0$. A dynamical map is unital if
\begin{equation} 
\Lambda(t)\mathbb{1}_{\rm S}=\mathbb{1}_{\rm S} \,,
\end{equation}
which requires $\vec{d}=\vec{0}$. These maps have the distinct property that they never increase the initial purity ($tr\rho_{\rm S}^{2}$) of any state. When restricting to qubit dynamical maps, they can always be Kraus decomposed using unitary Kraus operators. That is the reduced dynamics generated by unital channels may be constructed by averaging over unitary channels.

It is worth noting that for the 2-qubit model there is a connection between the free Hamiltonian and the magnitude of the shift. The magnitude of the shift is zero iff the state $\Omega_{\rm E}=\frac{1}{2}(\mathbb{1}_{\rm S}\otimes\rho_{\rm E}(0))$ and $H_{\rm free}$ are orthogonal operators i.e. Tr($\Omega_{\rm E}H_{\rm free}$)=0. But as we see in the next section, there are special choices of Hamiltonian parameters that make the dynamical map unital independent of the choice of initial environment state.

\subsection{Three-qubit model of dynamical symmetry breaking}

The reduced dynamics for the three qubit model is computed from
\begin{equation}
    \tilde{\Lambda}^{ab}(t,0)=\frac{1}{4}tr\left[\sigma^{a}_{\rm S}\otimes\mathbb{1}_{\rm E}\otimes\mathbb{1}_{\rm E'}\tilde{U}(t,0)\sigma^{b}_{\rm S}\otimes(\mathbb{1}_{\rm E }+z_{\rm E}(0)Z_{\rm E})\otimes\rho_{\rm E'}(0)\tilde{U}^{\dag}(t,0)\right]
\end{equation}
where we assume that the second qubit is in an initial state compatible with phase covariance. For this reason we do not include initial correlations between subsystems $E$ and $E^{\prime}$, although this is not necessary for the reduced dynamics to be CP. 

For $t<\tau$, we claim that the dynamical map is phase covariant. Using the results of the previous section we find
\begin{equation}
\tilde\Lambda_{t<\tau}(t)=\begin{bmatrix} 1&0&0&0\\0&\cos^{2} 2 \omega t&-\cos{2\omega t}\sin{2\omega t}&0 \\ 0&\cos{2\omega t}\sin{2\omega t}&\cos^{2}{2\omega t}&0 \\ z_{\rm E}\sin^{2}{2\omega t}&0&0&\cos^{2}{2\omega t}
\end{bmatrix}
\end{equation}
If we now focus on the unital action of $\tilde{\Lambda}_{t<\tau}$
\begin{equation}
    \tilde{T}_{t<\tau}(t)=\begin{bmatrix}\cos{2\omega t}&-\sin{2\omega t}&0 \\ \sin{2\omega t}&\cos{2\omega t}&0\\0&0&1\end{bmatrix}\begin{bmatrix}\cos{2\omega t}&0&0\\0&\cos{2\omega t}&0 \\0&0&\cos^{2}{2\omega t}\end{bmatrix}\,,
\end{equation}
 it is evident that any rotation about the $z$-axis will commute with $\tilde{T}_{t<\tau}(t)$. Since the shift is left fixed by rotations about the $z$-axis, it follows that
 \begin{equation}
     \tilde{\Lambda}_{t<\tau}(t)[e^{-i\omega t Z_{\rm S}}\rho_{\rm S}e^{i\omega t Z_{\rm S}}]=  e^{-i\omega t Z_{\rm S}}[\tilde{\Lambda}_{t<\tau}(t)\rho_{\rm S}]e^{i\omega tZ_{\rm S}}\, ,
 \end{equation}
 i.e. the reduced dynamics is phase covariant as it has a time-translation symmetry generated by $\omega Z_{\rm S}$.
 The phase-covariant map presented here is known as the generalized amplitude damping channel, which appears in quantum information to describe dissipative processes \cite{nielsen00}. 

For $t\geq \tau$, where we introduce the phase covariance breaking into the interaction, the dynamical map is found to be 
\begin{equation}
    \tilde{\Lambda}_{t\geq\tau}(t)=\begin{bmatrix}1&0&0&0\\\tilde{d}_{x}(t)&0&0&0\\\tilde{d}_{y}(t)&0&0&0\\\tilde{d}_{z}(t)&0&0&0\end{bmatrix}\,,
\end{equation}
where
\begin{equation}
    \begin{split}
    \tilde{d}_{x}(t)&=\frac{z_{\rm E}x_{\rm E'}}{\sqrt{2}}\Bigg[\bigg(g_{+}g_{-}-\Re{[b_{-}g_{+}]}-\Im{[g_{-}h_{-}]}+\Re{[a_{+}]}\Re{[h_{-}]}-\Re{[b_{-}]}\Im{[h_{-}]}\bigg) \\ &+\bigg(g_{+}g_{-}-\Im{[g_{+}h_{+}+g_{-}h_{-}]}+\Re{[h_{+}h_{-}}]\bigg)\cos{4\omega \tau}\\& +\bigg(\Im{[h_{+}h_{-}]}-\Re{[g_{+}h_{+}+g_{-}h_{-}]}\bigg)\sin{4 \omega \tau}\Bigg]\\
    \tilde{d}_{y}(t)&=\frac{z_{\rm E}x_{\rm E'}}{\sqrt{2}}\Bigg[\bigg(\Re{[g_{-}h_{-}]}-\Re{[a_{+}g_{+}]}+\Re{[b_{-}]}\Re{[h_{-}]}+\Re{[a_{+}]}\Im{[h_{-}]}\bigg)
    \\ &+\bigg(\Re{[g_{+}h_{+}]}+\Re{[g_{-}h_{-}]}+\Im{[h_{+}\bar{h}_{-}]}\bigg)\cos{4\omega \tau} \\&+\bigg(-g_{+}g_{-}+\Im{[g_{+}\bar{h}_{+}]}+\Im{[g_{-}h_{-}]}+\Re{[h_{+}h_{-}}]\bigg)\sin{4 \omega \tau}\Bigg]  \\
    \tilde{d}_{z}(t)&=\frac{z_{\rm E}}{2}
    \Bigg[\bigg(|a_{+}|^{2}-|b_{-}|^{2}+g^{2}_{-}+2\Im{[g_{+}h_{-}]}\bigg) +\bigg(g^{2}_{-}-|h_{+}|^{2}-2\Im{[g_{+}h_{-}]}\bigg)\cos{4\omega\tau} \\ &+2\bigg(\Re{[g_{-}h_{+}]}-\Re{[h_{-}]}\Im{[h_{-}]}\bigg)\sin{4\omega \tau}\Bigg] 
    \end{split}\,
\end{equation}
and $a_{\pm}$, $b_{\pm}$, $g_{\pm}$, and $h_{\pm}$ are defined in Eq.(\ref{eq:abgh}), and are all evaluated at $t-\tau$. The dynamical map after phase covariance breaking is a time-dependent SWAP operation, mapping all initial states to $\vec{\tilde{d}}(t)$. The transition of the reduced dynamics is continuous iff $\tau=\frac{\pi}{2\omega}(n+\frac{1}{2})$ for some positive integer $n$; i.e. only if $\tau$ coincides with a time when $\Lambda_{t<\tau}$ is non-invertible. If $\tau$ is not finely tuned, the image of  $\Lambda_{t<\tau}(\tau)$ will suddenly collapse to the point $\vec{\tilde{d}}(\tau)=z_{\rm E}(0)\hat{z}$.

\section{Features of the reduced dynamics}
\label{sec:features}
 This section reviews important non-unitary features of the dynamical map families that we have found in the previous section. We begin with a look at the invertibility of the dynamical maps, as this determines the structure of possible master equations used to generate the reduced dynamics. We also look at the classes of dynamical maps for the 2-qubit model, which we characterize by their degree of symmetry.


\subsection{Invertibility}
 Of paramount interest is the invertibility structure, which determines when time-local master equations are viable for use as generators. In the case of the 3-qubit example it is clear that the map is always non-invertible as $\tilde{T}_{t\geq\tau}(t)=0$. 
 
 But the invertibility of the dynamical map for the two-qubit is non-trivial, and we show in this section how it depends on the Hamiltonian and the state of the environment. We find that there are in fact three independent ingredients that play a role in defining the invertibility: the strength of the system/environment coupling, the ratio between eigenenergies, and the initial environment state. The fact that these are distinct criteria is seen from the special row structure of $\Lambda$ 
 \begin{equation}
    {\rm Det}\,\Lambda(\tau) = {\rm Det} \,\vec{\vec{T}}(\tau)= |\vec{T}_{z}(\tau)|^{2} 
\end{equation}
 where $\vec{T}_{z}$ is the third row of $\vec{\vec{T}}$. This now leads to three conditions that determine when the dynamical map is non-invertible as we have 
 \begin{equation}
       {\rm Det}\,\Lambda(\tau) = 0 \Leftrightarrow \vec{T}_{z}(\tau)= \vec{0}\,.
   \end{equation}
  
Starting with the $z$-component, $\Lambda^{zz}(\tau)=0$ implies
\begin{equation}
    \sin^2\phi_{+}\sin^2(\omega_{+}\tau)+\sin^2\phi_{-}\sin^2(\omega_{-}\tau)=1\,.
    \label{eq:TzBs}
\end{equation}

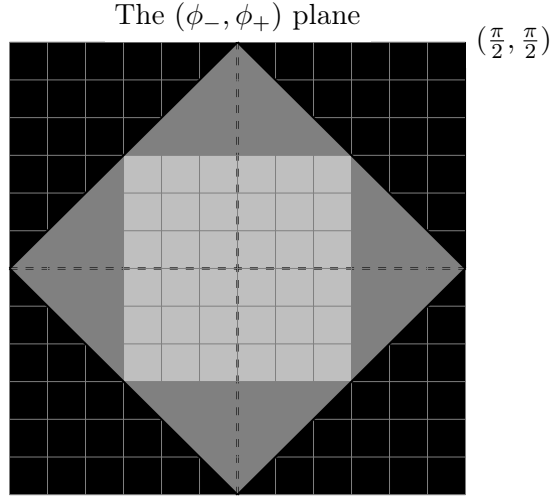
\begin{figure}[h]
\begin{adjustbox}{center}
\begin{tikzpicture}
\fill[black] (-3,0) -- (0,3) -- (-3,3); 
\fill[black] (-3,0) -- (0,-3) -- (-3,-3) ; 
\fill[black] (0,-3) -- (3,0) -- (3,-3); 
\fill[black] (3,0) -- (0,3) -- (3,3) ;
\fill[lightgray] (1.5,1.5) -- (1.5,-1.5) -- (-1.5,-1.5) -- (-1.5,1.5) ;
\fill[gray](-1.5,1.5)--(1.5,1.5)--(0,3) ;
\fill[gray](-1.5,1.5)--(-1.5,-1.5)--(-3,0) ;
\fill[gray](1.5,1.5)--(1.5,-1.5)--(3,0) ;
\fill[gray](-1.5,-1.5)--(1.5,-1.5)--(0,-3) ;
\draw[black,dashed,thick] (-3,0) -- (3,0) ;
\draw[black,dashed,thick] (0,-3) -- (0,3) ;
\draw[step=.5cm,gray, very thin] (-3.0,-3.0) grid (3.0,3.0);
\draw[black,thick] (-3.0,0.0) -- (0.0,3.0)node[anchor=south,fill=white]{The $(\phi_{-},\phi_{+})$ plane};
\draw[black,thin] (3,0)--(3,3) node[anchor=west,fill=white]{($\frac{\pi}{2},\frac{\pi}{2})$};
\draw[black,thick] (0,3) -- (3,0); 
\draw[black,thick] (3,0) -- (0,-3) ;
\draw[black,thick]  (0,-3) -- (-3,0);
\end{tikzpicture}
\end{adjustbox}
\caption{\raggedright \small The $(\phi_{+},\phi_{-})$ plane split into regions based on the non-local properties of the dynamics. The light gray region contains Hamiltonians that are not perfect entanglers, whereas the other regions (black+dark grey) are perfect entanglers. The entire gray region supports time local dynamics. The black region is the strong coupling region and contains the only Hamiltonians that may have ${\rm Det}\Lambda=0$.}
\label{fig:Non-Local}
\end{figure}

Eq.(\ref{eq:TzBs}) may only be satisfied at a discrete set of times, and only if $\sin^{2}\phi_{+}+\sin^{2}\phi_{-}\geq 1$. The set of $\phi_{\pm}$ that satisfy this condition constitute the strong coupling regime of parameter space. Figure \ref{fig:Non-Local} shows that this region of parameter space partially overlaps with the space where $U(t)$ is a perfect entangler.

However, strong coupling between system and environment is not a sufficient for the map to be non-invertible. Eq.(\ref{eq:TzBs}) also depends on the energy eigenvalues. For example consider the boundary between the dark gray and black regions of Figure \ref{fig:Non-Local} where $\sin^{2}\phi_{+}+\sin^{2}\phi_{-}=1$ (excluding the corners where either $\phi_{+}=0$ or $\phi_{-}=0$). In this region Eq.(\ref{eq:TzBs}) is satisfied iff, 
\begin{equation}
    \sin^{2}\omega_{+}\tau=\sin^{2}\omega_{-}\tau=1 \,.
\end{equation}
 In order for this to be possible integers $k$ and $l$ must exist such that, 
 \begin{equation}
         \frac{\omega_{+}}{\omega_{-}}=\frac{2k+1}{2l+1} \,.
 \end{equation}
 Defining $\nu$ such that $\omega_{+}=(2k+1)\nu$ and $\omega_{-}=(2l+1)\nu$; for any positive integer n a non-invertiblity can appear in the reduced dynamics at times given by, 
 \begin{equation}
     \tau_{n}=\frac{(2n+1)\pi}{2\nu}\,.
 \end{equation}
 So on this boundary only Hamiltonians with commensurate eigenenergies require time-non-local dynamics (although incommensurate Hamiltonians on the boundary can generate dynamical maps with small determinant). On the other hand, away from the boundary, farther into the blue region, most frequency pairs will generate dynamics with non-time-locality for some initial environment states. 
 
So far, we have considered only one of the conditions for non-invertibility, $\Lambda^{zz}(\tau)=0$, which imposes conditions on the Hamiltonian that must be satisfied for the map to be non-invertible. However, constraining the Hamiltonian to be in the strong-coupling regime is not sufficient: the initial environment state also plays a role. The class of initial states of the environment that lead to non-time-local dynamics is found by setting the remaining components of $\vec{T}_{z}(\tau)$ to zero, 
\begin{equation}
    \begin{split} 
        &\Lambda^{zx}(\tau)= \Lambda^{zxx}(\tau)x_{E}+\Lambda^{zxy}(\tau)y_{E} = 0\,, \\ 
        &\Lambda^{zy}(\tau)=\Lambda^{zyx}(\tau)x_{E}+\Lambda^{zyy}(\tau)y_{E} = 0 \,.
    \end{split}
\end{equation}
We can characterize the set of initial states for which non-invertibility will occur by finding the vector $\vec{\eta}(\tau)$ in the $xy$ plane associated to the environment qubit state for which both 
\begin{equation}
 V(\tau_i) \vec{\eta}(\tau)=
\begin{bmatrix}\Lambda^{zxx}(\tau) & \Lambda^{zxy}(\tau) \\ \Lambda^{zyx}(\tau) & \Lambda^{zyy}(\tau)\end{bmatrix}\vec{\eta}(\tau)=0\,
\end{equation}
and ${\rm Det}\,V(\tau)=0$ at some fixed $\tau$. The solution is 
\begin{equation}
    \vec{\eta}(\tau)=\Lambda^{zyy}(\tau)\hat{x}-\Lambda^{zyx}(\tau)\hat{y}\,,
\end{equation}
and the orthogonal direction is
\begin{equation}
    \vec{\eta}^{\perp}(\tau)= \Lambda^{zxx}(\tau)\hat{x}+\Lambda^{zxy}(\tau)\hat{y}\,.
    \label{eq:etaperp}
\end{equation}
A non-invertibility at time $\tau$ can be removed by shifting the initial state to contain a component in the direction $\vec{\eta}^{\perp}(\tau)$. Furthermore, there can only be a discrete set of times, $\tau_{i} < T$, where the condition $\Lambda^{zz}(\tau_{i})=0$ can be satisfied. Assume there are $N$ non-invertible times, 
\begin{equation}
    0 < \tau_{1}  < ... < \tau_{\rm N} < T, 
\end{equation}
with the associated initial environment states that preserve the non-invertibility $\{\hat{\eta}_{1},..., \hat{\eta}_{N}\}$. Single out $\hat{\eta}_{1}$ and note that so long as $\hat{\eta}^{\perp}_{1}\cdot\hat{\eta}^{\perp}_{k}\neq 0$, then the non-invertibility at $\tau_{k}$ is eliminated by the presence of $\hat{\eta}^{\perp}_{1}$ in $\vec{r}_{\rm E}(0)$. The remaining directions all must satisfy $\hat{\eta}^{\perp}_{1}\cdot \hat{\eta}^{\perp}_{k}=0$. Since these initial environment states lie in a two dimensional space, all the remaining $\hat{\eta}^{\perp}_{i}$ must equal $\hat{\eta}_{1}$. The question of removing all non-invertibilities up to $T$ is then equivalent to finding a $\theta$ such that, 
\begin{equation}
    \eta^{\perp}_{i} \cdot (\cos\theta\hat{\eta}^{\perp}_{1}+\sin\theta\hat{\eta}_{1})\neq 0 ,
    \label{eq:etaperpall}
\end{equation}
for all $i\in\{1, N\}$. 

We have demonstrated that certain off diagonal components (coherences) of the initial environment state in the eigenbasis of the free  environment Hamiltonian ($H_{\rm E;\rm free}$) control the appearance of time-non-locality in the reduced dynamics. In Section \ref{sec:mastereqn} we use this knowledge of the time-local environment states in order to construct exact non-local master equations as well as approximate time-local master equations. 

\subsection{Dynamical map families} 
We saw above how the invertibility of the dynamical map can be dramatically changed by symmetry breaking during the dynamics. On the other hand, for the time-independent Hamiltonian, non-invertibility only occured in a subset of the parameter space. In this section we classify all the dynamics possible with the Hamiltonian in Eq.(\ref{eq:H}). This classification illustrates the relationship between symmetry, non-invertibility, and two other key properties of open system dynamics: non-Markovianity and unitality. 

 Table \ref{table} lists the various dynamical map families contained in this model, together with each family's characteristics that we derive below. The largest family is the non-commensurate family ($\mathcal{N}$), with two independent frequencies. If we consider the time evolution in the Bloch ball, the trajectories generated are dense for any initial state $\vec{r}_{\rm S}(0)$. As mentioned previously, we use the (AB) frame to discuss the symmetries as they are simplest in this frame. For example the symmetries present in the entire family of Hamiltonians are $Z_{\rm A}\otimes \mathbb{1}_{\rm B}$ and $\mathbb{1}_{\rm A}\otimes Z_{\rm B}$. Of course these are equivalent to $H$ and $P_{zz}$, however we find that this frame compresses the discussion of Hamiltonian families that have additional symmetries beyond these two. 

The Hamiltonians which generate the family $\mathcal{D}$ have more symmetries than those corresponding to $\mathcal{N}$; these Hamiltonians commute with all rotations performed in the $B$ subsystem. The apparent symmetry between $A$ and $B$ in Eq.(\ref{eq:HAB}) is broken by the relationship to the $S$/$E$ frame. The presence of more symmetries simplifies the time-dependence of the dynamical map, thus less complicated trajectories are generated and the non-Markovian measures are periodic. The witnesses of non-Markovianity become periodic even in the case that the eigenfrequencies of the Hamiltonian are commensurate ($\mathcal{C}$). $\Lambda \in \mathcal{D}$ are not structurally different than those in $\mathcal{N}$ i.e. no additional components or partial components vanish. $\mathcal{D}$ does not have conserved quantities at the level of reduced dynamics.  

There are special subfamilies of $\mathcal{D}$ that support conserved quantities at the level of reduced dynamics. These families $\mathcal{D}_{\pm}$ contain dynamical maps that are phase damping channels, which are unital channel with the additional condition that $\lambda_{x}=1$ and $\lambda_{y}=\lambda_{z}=\lambda$. They are generated from the Hamiltonians with parameters such that $\phi_{+}=\pm\phi_{-}$. These families are simple to study and allow us to easily determine the set of initial environment states that yield invertible dynamics. For example in the family $\mathcal{D}_{+}$, there is only 1 distinct direction $\vec{\eta}(\tau)$ that allows non-invertibility namely $\vec{\eta}=\hat{y}$. 
 
 Restricting further the allowed set of initial environment states uncovers a set of Markovian families $\mathcal{M}_{\pm}$. These families contain the maps in $\mathcal{D}_{\pm}$ generated using $\vec{r}_{\rm E}(0)=\hat{x}$ ($\mathcal{M}_{+}$) and $\vec{r}_{\rm E}(0)=\hat{y}$ ($\mathcal{M}_{-}$). This is simply explained by looking at the Kraus decomposition of these channels. To illustrate this take an element of $\mathcal{D}_{+}$. The minimal Kraus decomposition has the form 
 \begin{equation}
     \Lambda_{+}\rho_{\rm S}=\frac{1+x_{\rm E}}{2}U_{+}\rho_{\rm S}U^{\dag}_{+}+\frac{1-x_{\rm E}}{2}(Z_{\rm S}U_{+}Z_{\rm S})\rho_{\rm S}(Z_{\rm S}U^{\dag}_{+}Z_{\rm S})
 \end{equation}
 thus when the initial environment state is pure and in the $x$-direction the channel becomes unitary.  We say the dynamics is Markovian as the dynamical map is unitary and thus generated by a time-dependent Lindblad equation with non-negative rates. Note that  unitary reduced dynamics is the only kind of Markovian dynamics possible, since the environment Hilbert space is of finite dimension, so necessarily has a free Hamiltonian that is bounded from below \cite{Wolf_2008}. As in all Markovian open systems, the initial environment state remains fixed under the time evolution in $\mathcal{M}_{\pm}$. 
 
So far we have not mentioned the family of maps that are phase covariant i.e. the generalized amplitude damping channels $\mathcal{A}_{\pm}$, often referred to by the acronym GADC. These families do not neatly fit within the previous families, but instead have non-zero overlap with $\mathcal{N}$, $\mathcal{C}$, and $\mathcal{D}$. These maps have singular values satisfying $\lambda_{z}=\lambda_{x}^{2}=\lambda^{2}_{y}$, and are non-unital with the only non-zero component given by $d_{z}=z_{\rm E}(1-\lambda_{z})$. For $z_{\rm E}=1$ the maps are simply known as amplitude damping channels. These maps are generated when either $\Delta_{-}=\kappa_{+}=0$ ($\mathcal{A}_{+}$) or $\Delta_{+}=\kappa_{-}=0$ ($\mathcal{A}_{-}$) and $\vec{r}_{\rm E}(0)=z_{\rm E}(0)\hat{z}$. 

It is worth noting that the families $\mathcal{D}_{\pm},\mathcal{A}_{\pm}$,and $\mathcal{M}_{\pm}$ consist entirely of extreme channels; i.e. channels that live in the boundary of the set of qubit channels\cite{Ruskai2001}. The extreme channels form a tetrahedron, and consist of minimal rank Kraus operators, where maps on the interior have non-minimal Kraus rank. The families $\mathcal{M}_{\pm}$ are contained in the corner representing the trivial channel. The families $\mathcal{D}_{\pm}$ live on an edge connected to the trivial channel. And the families $\mathcal{A}_{\pm}$ live on a face of the tetrahedron.

 \begin{center}
 \begin{tabular}{ |p{4.2cm}||p{4.0cm}|p{3.1cm}|p{2.0cm}|}
 \hline
 \multicolumn{4}{|c|}{Dynamical Map Families} \\
 \hline
 \quad \quad \quad \quad Family & \, Time Evolution & Extra Symmetries& \quad Unital  \\
 \hline
 Non-Commensurate ($\mathcal{N}$) \newline $\omega_{+}\neq  q\omega_{-}$ &  Aperiodic \newline Non-Markovian & \quad \quad \quad None &  If $z_{\rm E}(0)=0$ \\
 \hline 
 Commensurate ($\mathcal{C}$) \newline $\omega_{+}=q\omega_{-}$&Periodic \newline Non-Markovian&\quad \quad \quad None & If $z_{\rm E}(0)=0$  \\
 \hline 
  Degenerate ($\mathcal{D}$) \newline $\omega_{+}=\omega_{-}$& Periodic \newline \quad Non-Markovian &$\{\mathbb{1}_{\rm A}\otimes X_{\rm B}$, $\mathbb{1}_{\rm A}\otimes Y_{\rm B}\}$& If $z_{\rm E}(0)=0$\\
 \hline
 Phase Damping($\mathcal{D}_{\pm}$) \newline $\phi_{+}=\pm \phi_{-}$& Periodic \newline Non-Markovian & $\{\mathbb{1}_{\rm A}\otimes X_{\rm B}$, $\mathbb{1}_{\rm A}\otimes Y_{\rm B}\}$ &  Any $\vec{r}_{\rm E}(0)$ \\
 \hline 
 Markovian ($\mathcal{M}_{\pm}$) \newline $\vec{r}_{\rm E}(0)$ = $\hat{x}$(+) or $\hat{y}$(-)& Markovian& $\{\mathbb{1}_{\rm A}\otimes X_{\rm B}$, $\mathbb{1}_{\rm A}\otimes Y_{\rm B}$\} &  Always \\ 
 \hline 
 Amplitude Damping \newline ($\mathcal{A}_{\pm}$)& Aperiodic NM ($\mathcal{A}_{\pm}\cap\mathcal{N}$) \newline Periodic NM \,\,\,($\mathcal{A}_{\pm}\cap \mathcal{D}$)  & \quad \quad \quad None \newline $\{\mathbb{1}_{\rm A}\otimes X_{\rm B}$, $\mathbb{1}_{\rm A}\otimes Y_{\rm B}$\} & If $z_{\rm E}(0)=0$\\ \hline
 \end{tabular} 
 \label{table}
 \end{center}
The method we devise later for constructing effective master equations relies on there being a set of fundamental time scales that can be used to determine all non-invertible times of the reduced dynamics. Such a task is especially simple for the families $\mathcal{D}_{\pm}$ and $\mathcal{A}_{\pm}\cap\mathcal{D}$, where only one such time scale exists. But for the set of Hamiltonians where the parameters are not carefully chosen ($\mathcal{N}$), such a construction is unfavorable as the non-invertible times behave as essentially random.

\subsection{Entanglement generation and non-Markovianity}
For the range of parameters possible in the two-qubit Hamiltonian, we can characterize the flow of information between system and environment by examining the non-Markovianty and the degree to which the system and environment can become entangled. 

The parameters required for the time-evolution to be (periodically) perfectly entangling can be found using the criteria of Makhlin \cite{Makhlin:2000, Zhang:2003zz}, that the convex hull of the eigenvalues of the matrix $m(U)=(Q^{\dagger}UQ)^{\rm T}Q^{\dagger}UQ$ contains zero, where $Q$ is the operator that changes to the Bell basis. Evaluating the eigenvalues of $m(U)$, the convex hull condition becomes (see details in Appendix \ref{appendix:entangle})
\begin{equation}
     \cos^{2}\varphi \beta_{+}^{2}(t)+\sin^{2}\varphi \beta_{-}^{2}(t)=\frac{1}{2}\,,
 \end{equation}
where $\varphi\in[0,2\pi)$ parameterizes the remaining convex combinations. This can be satisfied iff the largest of $\beta^{2}_{+}(t)$ and $\beta^{2}_{-}(t)$ is greater than or equal to $\frac{1}{2}$, which will hold at some times as long as ${\rm Max}(\phi_{+},\phi_{-})\geq \frac{\pi}{4}$. This condition on the parameter space of the Hamiltonian is shown in Figure \ref{fig:Non-Local}, compared with other conditions we derive below related to properties of the reduced dynamics. 

For nearly all parameter values in the Hamiltonian, Eq.(\ref{eq:HSEblocks}), and initial states, the dynamical map is non-Markovian. This is expected since the system and environment are the same (small) size. The  non-Markovianity is diagnosed by information back-flow into the system from the environment \cite{Laine2010}, with standard indicators being non-monotonicity in the evolution of trace distance and fidelity. For any two states on the reduced system, $\vec{r}_{1}(t)$ and $\vec{r}_{2}(t)$, the trace distance ($\mathcal{D}$) and fidelity ($\mathcal{F}$) are \cite{Chen2002},
\begin{equation}
 \begin{split}
     &2\mathcal{D}(\vec{r}_{1}(t),\vec{r}_{2}(t))= ||\vec{r}_{1}(t)-\vec{r}_{2}(t)|| \\ 
     &2\mathcal{F}(\vec{r}_{1}(t),\vec{r}_{2}(t))=1+\vec{r}_{1}(t)\cdot\vec{r}_{2}(t)+\sqrt{(1-r^{2}_{1}(t))(1-r^{2}_{2}(t))} \,.
 \end{split}
 \end{equation}
 Generically these measures are oscillatory, and aperiodic unless $\omega_{+}=q\omega_{-}$ for some $q \in \mathbb{Q}$. 

\begin{figure}[H]
\begin{subfigure}{.45\linewidth}
  \centering
 \includegraphics[width=0.9\linewidth]{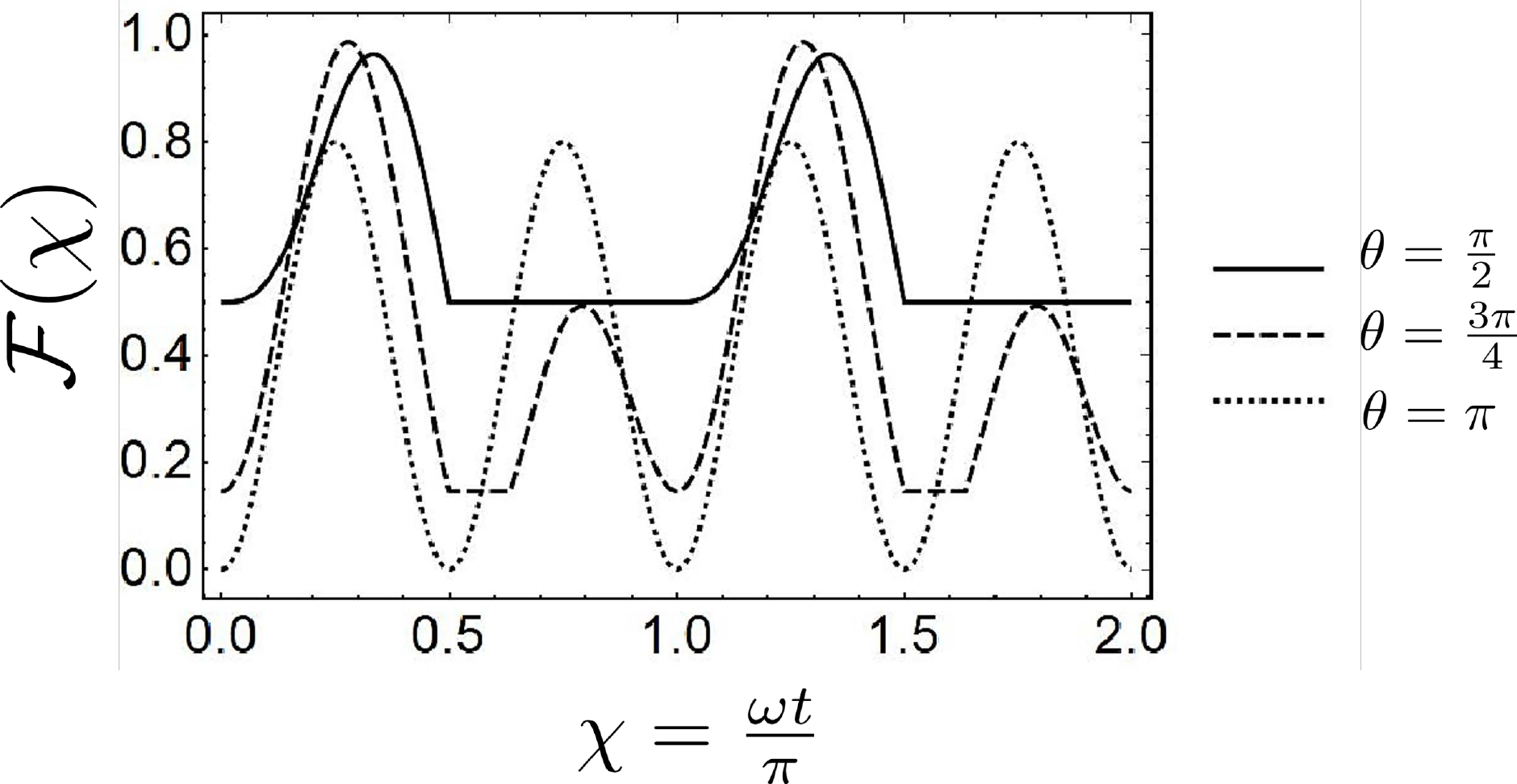}  
  \label{fig:FidePlot}
\end{subfigure} 
\begin{subfigure}{.45\linewidth}
  \centering
 \includegraphics[width=0.9\linewidth]{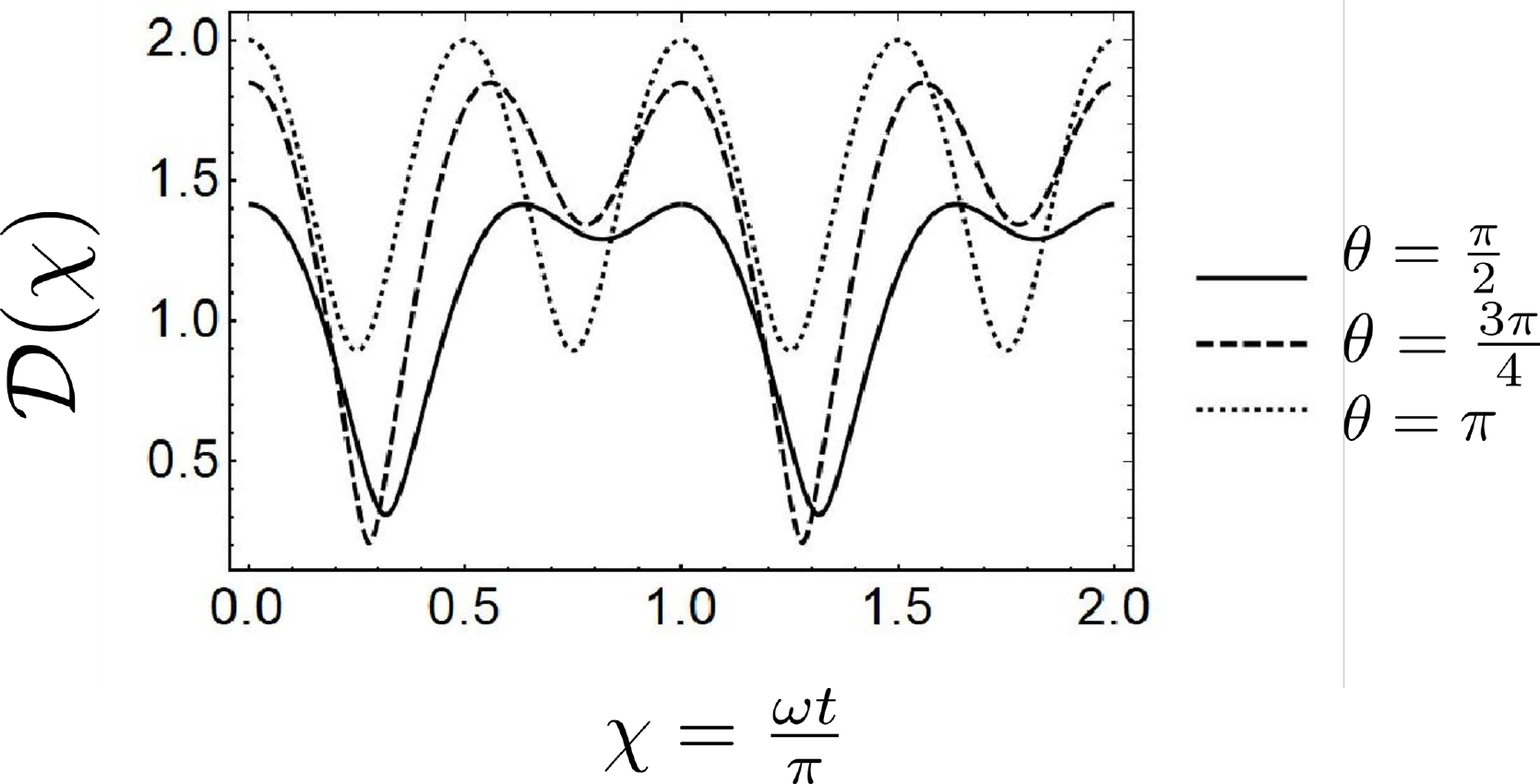}  
 \label{fig:DPlot}
\end{subfigure}
\caption{The fidelity and trace distance between time evolved reduced states using the initial conditions $\vec{r}_{1}(0)=\hat{x}$ and $\vec{r}_{2}(0)=\cos\theta\hat{x}+\sin\theta\hat{y}$ are plotted. The parameters used are $\omega=\omega_{+}=\omega_{-}=\frac{\sqrt{5}}{2}$,  $\tan\phi_{+}=\tan\phi_{-}=2$, and $\vec{r}_{\rm E}(0)=\vec{0}$. The non-Markovian nature of the reduced dynamics is evident from the oscillations of $\mathcal{D}$ and $\mathcal{F}$.}
\label{fig:fig}
\end{figure}

The degenerate family, Eq.(\ref{eq:degenerateDef}), always has periodic measures of non-Markovianity and is a useful case to look at in more detail. Figure \ref{fig:fig} shows the trace distance and fidelity in the degenerate family, demonstrating the non-Markovian character of the reduced dynamics. The dimensionless parameter $\chi=\frac{\omega t}{\pi}$ ($\omega=\omega_{+}=\omega_{-}$) is used to construct these plots, and the environment memory time-scale can be read off as $\tau_{\rm NM}\sim$ $\mathcal{O}(\frac{\pi}{2\omega})$.

Figure \ref{fig:fig} is generated with $\phi_{+}=\phi_{-}$ allowing for interesting features to appear in $\mathcal{F}$. If $\phi_{\pm}$ are perturbed so that $\phi_{+}\neq\phi_{-}$, these interesting features also vanish. First note the plateaus, which indicate there is a unitary phase (i.e. $\mathcal{F}$=constant) if we further require that $\vec{r}_{\rm E}(0)=\hat{x}$. Additionally, preceding the plateaus are discontinuities in $\dot{\mathcal{F}}$. In some other finite open systems, such discontinuities indicate dynamical phase transitions \cite{PhysRevB.102.220302}. 

Non-Markovianity of the dynamics has been equated to the indivisibility of the dynamical map into channels (CPTP maps) \cite{Laine2010}, however the actual relationship is more complicated as non-Markovian dynamical maps can be CP divisible \cite{2019A,2019B,2019C}. 
To study the divisibility of the dynamical map one looks at the interweaving maps ($\Phi$), defined using two times $\tau_{2}>\tau_{1}\geq0$, 
\begin{equation}
\Lambda(\tau_{2})=\Phi(\tau_{2},\tau_{1})\Lambda(\tau_{1}) \, .
\label{eq:interweave}
\end{equation}
For invertible reduced dynamics the interweaving map is computed as \newline $\Phi(\tau_{2},\tau_{1})=\Lambda(\tau_{2})\Lambda^{-1}(\tau_{1})$, although $\Phi$ may still be defined even when $\Lambda^{-1}$ does not exist \cite{2019B}. 

The degree of positivity of $\Phi(\tau_{2},\tau_{1})$ determines the divisibility class of the map $\Lambda(\tau_{2})$, as $\Lambda(\tau_{1})$ is completely positive by construction. It is important to note that divisibility is meant in a holistic sense i.e. the reduced dynamics is considered CP divisible up to time $\tau_{2}$ only if $\Phi(\tau_{2},\tau_{1})$ is completely positive for all $\tau_{1}<\tau_{2}$. A qubit channel is (infinitesimally) P-divisible if ${\rm Det}\Lambda(t) \geq 0$ for all $t\geq0$ \cite{Wolf_2008}. We have shown this to be true for the 2-qubit family studied. So the dynamical maps presented here may be arbitrarily split into positive maps, each made close to the identity channel. 

The question of CP divisibility is in general more subtle. For qubit dynamical maps of full rank, these maps are CP divisible if $s_{1}s_{2}s_{3}>0$, where $s_{i}$ are the singular values of the Lorentz normal form of the channel \cite{2019C}. For unital channels, this is equivalent to requiring ${\rm Det}\Lambda(t) >0$. We are able to find many examples of such maps. Take $\Lambda(t)\in \mathcal{N}$ and $z_{\rm E}(0)=0$. As long as the initial environment state is not pure, $\Lambda(t)$ will be a full Kraus rank unital channel. Therefore it will be CP divisible iff it is invertible. Of course these maps generate oscillation in non-Markovian measures, so do not constitute Markovian reduced dynamics. In Appendix \ref{appendix:divisibility}, we study the CP divisibility of deficient Kraus rank maps in $\mathcal{D}_{+}$. We find that even when these maps are invertible, there typically exist time intervals s.t. for $\tau \in [\tau_{a},\tau_{b}]$ the map $\Phi(\tau_{2},\tau)$ fails to be CP.   

\section{Master equations}
\label{sec:mastereqn}
We present the standard master equations that can be associated to any $\Lambda(t,0)$, both local and non-local in time. We find that by changing $\vec{r}_{\rm E}(0)$, different partitions of the singularity between non-local and local terms can be achieved. We show that we can capture some non-unitarity features of the reduced dynamics just using the local term.

\subsection{Standard master equations}

The dynamical map can be used to construct the time-local, although possibly singular, generator for the master equation 
\begin{equation}
    \partial_{t}\rho_{S}(t) = K_{\rm TL}(t)\rho_{S}(t)= \dot{\Lambda}(t)\Lambda^{-1}(t)\rho_{S}(t) \, .
\end{equation}
Since the inverse dynamical map is 
\begin{equation}
\begin{split}
    \Lambda^{-1}(t)& = \frac{1}{T^{2}_{z}}\begin{bmatrix} T^{2}_{z} &0&0&0\\ -d_{z} \vec{T}_{z}  & \vec{T}_{y} \times\vec{T}_{z} &\vec{T}_{z} \times\vec{T}_{x} &\vec{T}_{z} \end{bmatrix} 
\end{split} \, , 
\end{equation} 
the time-local generator is expressed in terms of the dynamical map components as
\begin{equation}
    \begin{split}
        K_{\rm TL}(t) = \frac{1}{T^{2}_{z}}\begin{bmatrix} 0&0&0&0 \\ -d_{z} (\dot{\vec{T}}_{x}\cdot \vec{T}_{z})& \dot{\vec{T}}_{x}\cdot(\vec{T}_{y}\times\vec{T}_{z}) &\dot{\vec{T}}_{x}\cdot(\vec{T}_{z}\times\vec{T}_{x})&\dot{\vec{T}}_{x}\cdot(\vec{T}_{z}) \\ -d_{z} (\dot{\vec{T}}_{y}\cdot \vec{T}_{z})& \dot{\vec{T}}_{y}\cdot(\vec{T}_{y}\times\vec{T}_{z}) &\dot{\vec{T}}_{y}\cdot(\vec{T}_{z}\times\vec{T}_{x})&\dot{\vec{T}}_{y}\cdot(\vec{T}_{z}) \\ -d_{z} (\dot{\vec{T}}_{z}\cdot \vec{T}_{z})& \dot{\vec{T}}_{z}\cdot(\vec{T}_{y}\times\vec{T}_{z}) &\dot{\vec{T}}_{z}\cdot(\vec{T}_{z}\times\vec{T}_{x})&\dot{\vec{T}}_{z}\cdot(\vec{T}_{z}) \end{bmatrix} \,.
    \end{split}
\end{equation}

\begin{figure}[H]
\begin{subfigure}{.48\textwidth}
  \centering
   \includegraphics[width=0.9\linewidth]{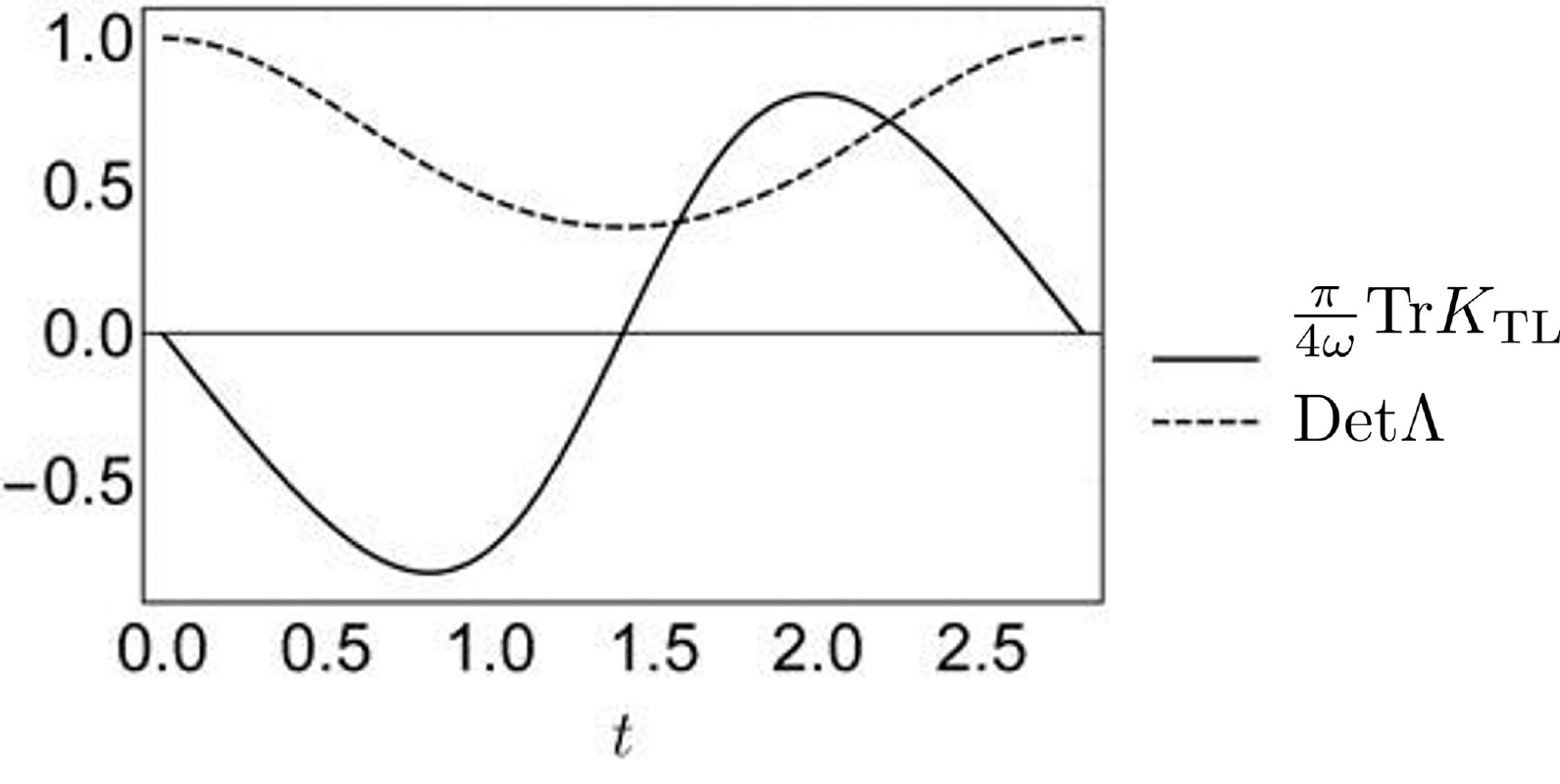}  \caption{Time-local reduced dynamics}
  \label{fig:Weak}
\end{subfigure}
\begin{subfigure}{.48\linewidth}
  \centering
 \includegraphics[width=0.9\linewidth]{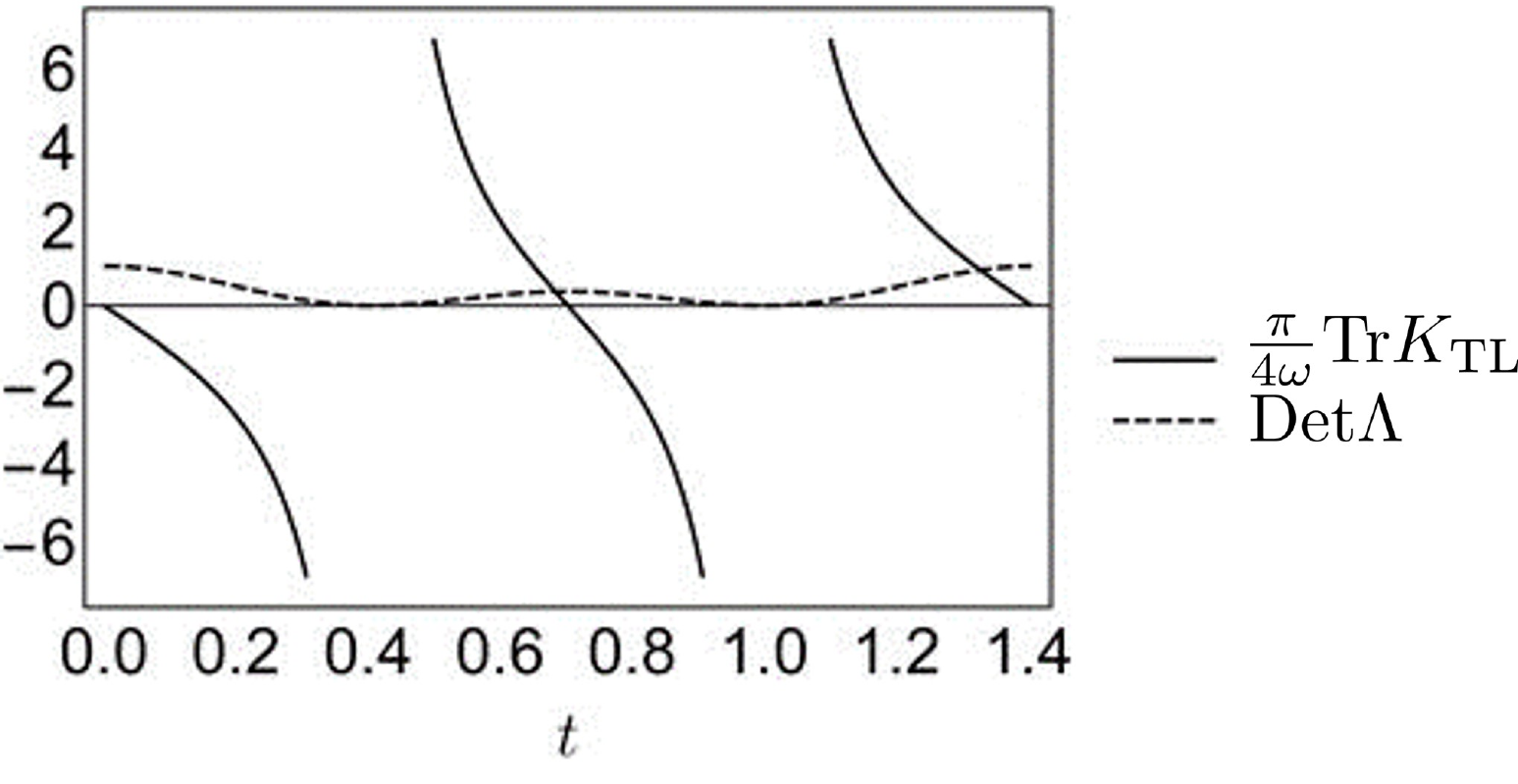}  \caption{Non-time-local reduced dynamics}
  \label{fig:Strong}
\end{subfigure}
\caption{ \small Tr$K_{\rm TL}$ (scaled by $\frac{\pi}{4\omega}$) and Det$\Lambda$ are plotted with $\vec{r}_{\rm E}(0)=\vec{0}$. a) Uses parameters $\omega=\omega_{+}=\omega_{-}=\frac{\sqrt{17}}{4}$ and $\tan\phi_{+}=\tan\phi_{-}=\frac{1}{2}$; the dynamics is invertible with Det$\Lambda$ separated from zero. b) Uses parameters $\omega=\omega_{+}=\omega_{-}=\sqrt{5}$ and $\tan\phi_{+}=\tan\phi_{-}=4$; Det$\Lambda$=0 has solutions, where the dynamical map becomes a projection onto the $\hat{Q}_{1}(\tau)$ direction in the Bloch ball.}
\label{fig:det}
\end{figure}

Fig (\ref{fig:det}) contains plots of ${\rm Det}\Lambda$ and ${\rm Tr}K_{\rm TL}$ using $\Lambda(t)\in \mathcal{D}_{+}$, which demonstrate that singularities in $K_{\rm TL}$ occur where ${\rm Det}\Lambda=0$. 
 
The generator $K_{\rm TL}$ can be put into Lindblad form, 
\begin{equation}
    \partial_{t}\rho_{S}(t)= -i[(\vec{H}_{\rm eff}(t)\cdot\vec{\sigma}_{\rm S}),\rho_{\rm S}(t)]+\sum_{i=1}^{3}\sum_{j=1}^{3}\gamma_{ij}(t)(\sigma^{j}_{\rm S}\rho_{\rm S}(t)\sigma^{i}_{\rm S}-\frac{1}{2}\{\sigma^{i}_{\rm S}\sigma^{j}_{\rm S},\rho_{\rm S}(t)\}) \, ,
\end{equation}
 where $\vec{H}_{\rm eff}(t)=\omega_{\rm S}\hat{z}+\vec{H}_{\rm open}(t)$ generates unitary evolution on $\mathcal{Q}_{\rm S}$ with a contribution from internal parameters and a portion that knows about the environment and interaction, $\vec{H}_{\rm open}(t)$. The coefficients $\gamma_{ij}(t)$ are the Lindblad coefficients which generate the non-unitarity that appears in the reduced dynamics. 
 
 \begin{figure}[H]
\begin{subfigure}{.48\textwidth}
  \centering
   \includegraphics[width=\linewidth]{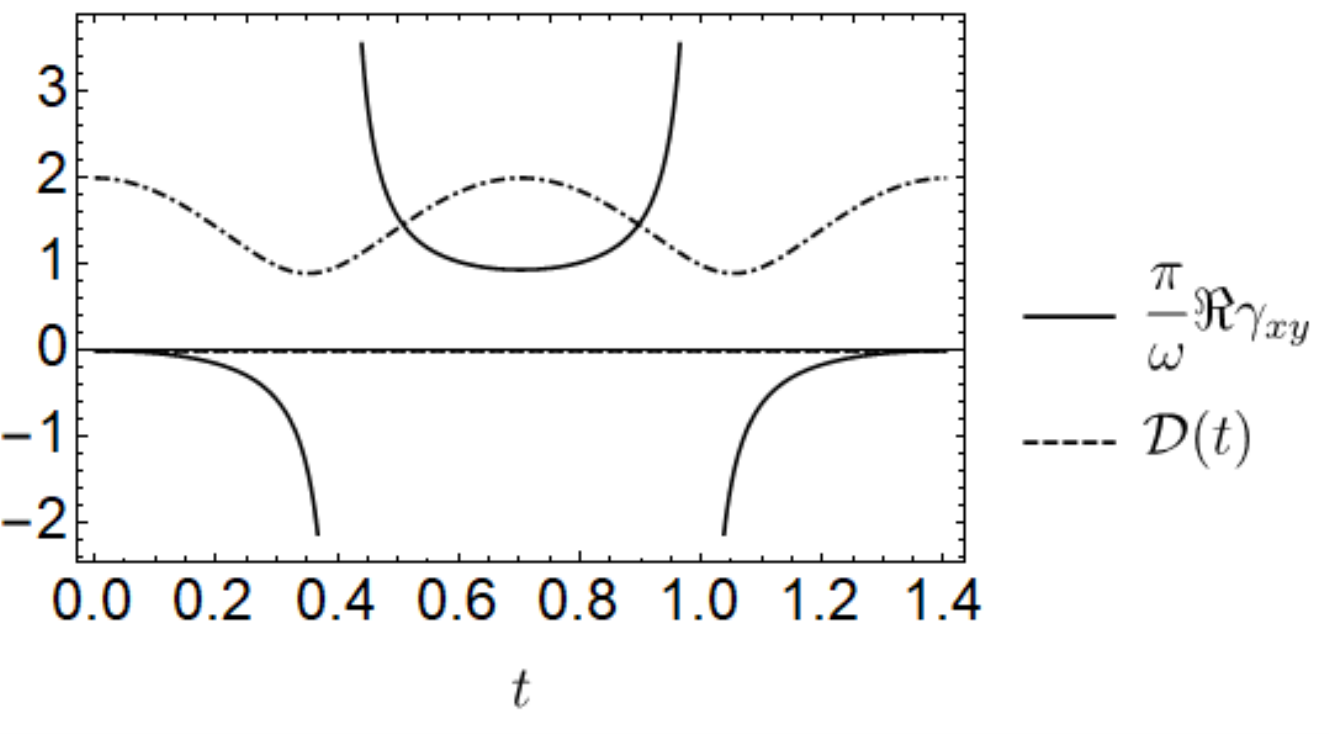}  
   \caption{$\Re \gamma_{xy}=\Re\gamma_{yz}$}
  \label{fig:Regamma}
\end{subfigure}
\begin{subfigure}{.48\linewidth}
  \centering
 \includegraphics[width=\linewidth]{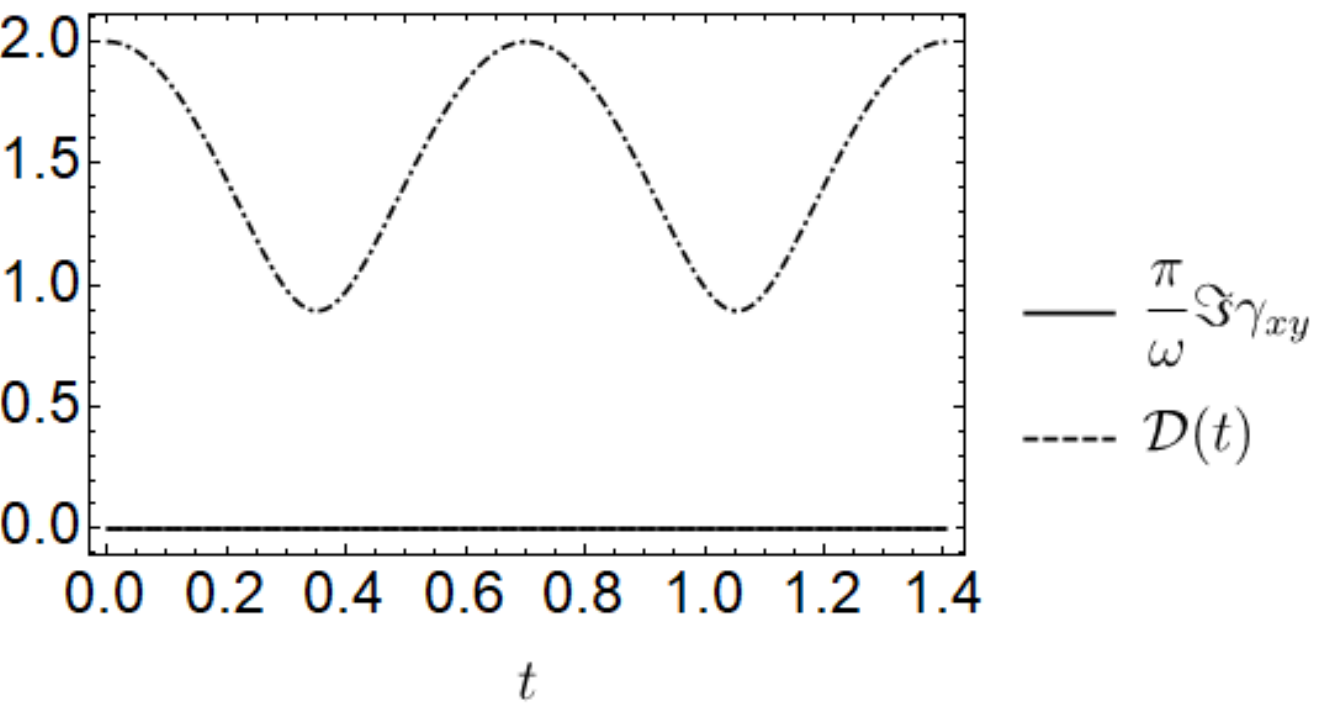}  \caption{$\Im\gamma_{xy}=\Im\gamma_{yz}$}
  \label{fig:Imgamma}
\end{subfigure}
\caption{\small Plotted are the real and imaginary parts of the off diagonal Lindblad coefficients. Plots are generated using the parameters $\omega_{+}=\omega_{-}=\frac{\sqrt{5}}{2}$ and $\tan\phi_{+}=\tan\phi_{-}=2$. Included in the plots is the trace distance evaluated using the same parameters and initial states $\vec{r}_{1}(0)=-\vec{r}_{2}(0)=\hat{x}$. We include the trace distance to demonstrate that experimentally accessible quantum information measures can constrain the parameters that appear in effective master equations; in this case oscillation frequencies of the Lindblad coefficients.}
\label{fig:gamma}
\end{figure}
 
 Obtaining the Lindblad form of the master equation amounts to a change of basis, for example the effective Hamiltonian has components 
 
 \begin{equation}
\begin{split}
    &H^{i}_{\rm eff}(t)=\frac{1}{2}\epsilon^{ijk}K_{jk}(t)
\end{split} \,,
\end{equation} 
and similar expression exist for the $\gamma_{ij}(t)$, though we do not include them here. Figure \ref{fig:gamma} shows how we can use the quantities such as the $\mathcal{D}(t)$ to determine the oscillation timescales that appear in the Lindblad coefficients, which is useful to set a scale for environment memory timescales that will inevitably appear in effective master equations.

In the strong coupling region where the time-local description has singular behavior, we instead use an integral master equation known as the Nakajima-Zwanzig equation. The Hamiltonian being time independent allows the NZ generator to take the form $K_{NZ}(t,\tau)=K_{NZ}(t-\tau)$. Thus the NZ equation is a convolution, 
\begin{equation}
    \partial_{t}\rho_{\rm S}(t)= \int_{0}^{t}K_{\rm NZ}(t-\tau)\rho_{\rm S}(\tau)d\tau \,.
\end{equation}
The convolution kernel may be obtained from the dynamical map using the Laplace transform of the dynamical map $\Phi(s)$ (as in \cite{Smirne2010}), 
\begin{equation}
    \tilde{K}_{\rm NZ}(s)= s\mathbb{1}_{4}-\Phi^{-1}(s)
\end{equation}
The form of the functions $\alpha_{\pm}(t)$ and $\beta_{\pm}(t)$ restrict the dynamical map components to having few Fourier components. That is each $\Lambda^{ab}$ has a Fourier decomposition of the form, 
\begin{equation}
\begin{split}
    \Lambda^{ab}(t)=C^{ab}+&F^{ab}_{1}e^{-i2\omega_{+}t}+F^{ab}_{2}e^{-i2\omega_{-}t}+F^{ab}_{3}e^{-i(\omega_{+}+\omega_{-})t}+F^{ab}_{4}e^{-i(\omega_{+}-\omega_{-})t} \\ +&G^{ab}_{1}e^{i2\omega_{+}t}+G^{ab}_{2}e^{i2\omega_{-}t}+G^{ab}_{3}e^{i(\omega_{+}+\omega_{-})t}+G^{ab}_{4}e^{i(\omega_{+}-\omega_{-})t}
\end{split}\,.
\end{equation}
If $ab$ is even (odd) then the component $\Lambda^{ab}$ involves only even (odd) time dependent functions. It is now simple to go to the frequency domain, where we can write the generic Laplace transformed components,
\begin{equation}
   \begin{split}
    \Phi^{ab}(s)=\frac{C^{ab}}{s}+&\frac{F^{ab}_{1}}{s+i2\omega_{+}}+\frac{F^{ab}_{2}}{s+i2\omega_{-}}+\frac{F^{ab}_{3}}{s+i(\omega_{+}+\omega_{-})}+\frac{F^{ab}_{4}}{s+i(\omega_{+}-\omega_{-})} \\ +&\frac{G^{ab}_{1}}{s-i2\omega_{+}}+\frac{G^{ab}_{2}}{s-i2\omega_{-}}+\frac{G^{ab}_{3}}{s-i(\omega_{+}+\omega_{-})}+\frac{G^{ab}_{4}}{s-i(\omega_{+}-\omega_{-})} 
\end{split} \,.
\end{equation}
In principle one can use these formulae to  construct the exact non-time-local master equation associated to the reduced dynamics, although the expressions involved are cumbersome. Instead for what follows we devise a method to expand the NZ kernel,
\begin{equation}
    K_{\rm NZ}(t-\tau) =\delta(t-\tau)K_{\rm TL}(t)+K_{\rm NTL}(t,\tau) \,.
\end{equation}
Such an expansion is available for each time-local initial environment state ($\vec{r}_{\rm TL}$), where the relative importance of the time-local vs non-time-local component is controlled by the magnitude of $\vec{r}_{\rm TL}-\vec{r}_{\rm E}(0)$.
\subsection{Effective master equation}

An exact time local master equation without singularities exists as long as either of the following conditions are met 
\begin{enumerate}[I]
    \item The Hamiltonian parameters are in the weak coupling region, i.e. $\sin^{2}\phi_{+}+\sin^{2}\phi_{-}<1$.
    \item $\vec{r}_{\rm E}$(0) has off diagonal elements in the eigenbasis of $H_{\rm E;\rm free}$ along certain directions described in Section IIIB.  
\end{enumerate}
 However, when neither of the above holds only a non-local equation can capture the exact time evolution of the reduced state. Such a master equation can be found using reduced dynamics from the time-local region.

Consider the case where the Hamiltonian parameters are in the strong coupling region and $\vec{r}_{\rm E}(0)=\vec{r}_{\rm NTL}$, yielding non-invertibilities in the reduced dynamics. By choosing a shifted initial environment state ($\vec{r}^{\,'}_{\rm E}$), we can engineer the time local component of a non-time local master equation. By using the exact non-time local dynamical map we can also determine the corresponding non-time-local component for the master equation. As the non-time-local piece is linear in $\delta\vec{r}=\vec{r}_{\rm NTL}-\vec{r}^{\,'}_{\rm E}$, the choice of shifted initial state controls its relevance. 

We can choose $\vec{r}^{\,'}_{\rm E}$ by considering states near to $\vec{r}_{\rm E}(0)$ that are shifted along the direction(s)
$\eta^{\perp}$ as defined in Eq.(\ref{eq:etaperp}). The time derivative of $\rho_{\rm S}(t;\vec{r}_{\rm NTL})$ may be expanded as,
\begin{equation}
\begin{split}
    \dot{\rho}_{\rm S}(t;\vec{r}_{\rm NTL})&=\partial_{t}\rho_{\rm S}(t;\vec{r}^{\,'}_{\rm E}+\delta \vec{r}) =\partial_{t}\left[\Lambda(t; \vec{r}^{\,'}_{\rm E})+(\Lambda(t;\delta\vec{r})-\Lambda(t;\vec{0}))\right] \rho_{\rm S}(0) \\
    &=K_{\rm TL}(t; \vec{r}^{\,'}_{\rm E})\rho_{\rm S}(t;\vec{r}^{\,'}_{\rm E})+\delta \vec{r}\cdot\left[ \frac{\partial\dot{\Lambda}(t;\delta \vec{r})}{\partial \delta \vec{r}}\right]\rho_{\rm S}(0) \\ 
    &= K_{\rm TL}(t; \vec{r}^{\,'}_{\rm E})\rho_{\rm S}(t;\vec{r}_{\rm NTL}) +\delta\vec{r}\cdot \left[\frac{\partial\dot{\Lambda}(t;\delta \vec{r})}{\partial \delta \vec{r}}-K_{\rm TL}(t;\vec{r}^{\,'}_{\rm E})\frac{\partial\Lambda(t;\delta \vec{r})}{\partial \delta\vec{r}}\right]\rho_{\rm S}(0) \,. 
\end{split}
\end{equation}
Note that since the dynamical map is linear in the initial environment state, we can replace $\frac{\partial\Lambda(t;\delta \vec{r})}{\partial \delta \vec{r}}$ in the above equation with $\frac{\partial\Lambda(t;\vec{r}^{\,'}_{\rm E})}{\partial \vec{r}^{\,'}_{\rm E}}$
\begin{equation}
\begin{split} 
   \dot{\rho}_{\rm S}(t;\vec{r}_{\rm NTL}) &= K_{\rm TL}(t; \vec{r}^{\,'}_{\rm E})\rho_{\rm S}(t;\vec{r}_{\rm NTL}) +\delta\vec{r}\cdot \left[\frac{\partial\dot{\Lambda}(t;\vec{r}^{\,'}_{\rm E})}{\partial \vec{r}^{\,'}_{\rm E}}-K_{\rm TL}(t;\vec{r}^{\,'}_{\rm E})\frac{\partial\Lambda(t;\vec{r}^{\,'}_{\rm E})}{\partial\vec{r}^{\,'}_{\rm E}}\right]\rho_{\rm S}(0) \,.
 \end{split}
 \end{equation}
This can be further simplified noting that $\dot{\Lambda}(t;\vec{r}^{\,'}_{\rm E})-K_{\rm TL}(t;\vec{r}^{\,'}_{\rm E})\Lambda(t;\vec{r}^{\,'}_{\rm E})=0$, yielding the final expression
 \begin{equation}
  \begin{split}
   \dot{\rho}_{\rm S}(t;\vec{r}_{\rm NTL}) &=K_{\rm TL}(t;\vec{r}^{\,'}_{\rm E})\rho_{\rm S}(t;\vec{r}_{\rm NTL})+\delta \vec{r} \cdot\left[\frac{\partial K_{\rm TL}(t;\vec{r}^{\,'}_{\rm E})}{\partial\vec{r}^{\,'}_{\rm E}}\Lambda(t;\vec{r}^{\,'}_{\rm E})\right]\rho_{\rm S}(0)  \,. 
  \end{split}
  \end{equation}
 We have now explicitly isolated a time-local component which is evaluated at the same Hamiltonian parameters as $\Lambda(t;\vec{r}_{\rm TNL})$, but uses the initial environment state $\vec{r}^{\,'}_{\rm E}$ instead of $\vec{r}_{\rm NTL}$. How large $\delta\vec{r}$ is compared with $\vec{r}_{\rm TNL}$ determines the maximum value of ${\rm Tr}K_{\rm TL}(t;\vec{r}^{\,'}_{\rm E})$, which in turn determines how relevant the integral term is in capturing the non-invertibility.

An effective time local description is obtained keeping only the zeroth order term, 
\begin{equation}
    \partial_{t}\rho_{S}(t;\vec{r}_{\rm NTL})\approx K_{\rm TL}(t;\vec{r}^{\,'}_{\rm E})\rho_{S}(t;\vec{r}_{\rm NTL})\,
\end{equation}
  Using this equation to generate the reduced dynamics is equivalent to exchanging the dynamical map $\Lambda(t,\vec{r}_{\rm NTL})$ for the dynamical map $\Lambda(t;\vec{r}^{\,'}_{\rm E})$, so obviously defines a completely positivity master equation.

 \section{Conclusions}
 \label{sec:conclude}
For a two-qubit system with symmetry, we have classified how the non-Markovianity, unitality, and time non-locality depend on the system/environment Hamiltonian and the initial state of the environment. We also considered a three-qubit example with a time-dependent interaction that broke a dynamical symmetry. 

The oscillatory character of the reduced dynamics demonstrates that there is a continuous flow of information flow back and forth between system and environment.  For sufficiently symmetric Hamiltonians, special initial environment states exist that support Markovian reduced dynamics, where the information flow becomes one-way. While this is not surprising, it is a reminder that the assumption of a Markovian master equations is not consistent with typical interacting Hamiltonians. This is true in the cosmological context as well \cite{Shandera_2018, Burgess:2022nwu}.

In typical effective theories of open systems, then, we need some guidance for how to parameterize the possible non-Makovian behavior. In the qubit model, whether the flow of information between system and environment was periodic or aperiodic depended on the degree of symmetry present in the Hamiltonian. There was no reason to assume that any of the parameters in the Hamiltonian take related values (for example, commensurate frequencies in the system/environment free Hamiltonians). In some cosmological contexts, however, there are effects that connect parameter values. For example, in inflation all modes undergo a common squeezing driven by the background expansion. This squeezing can determine a significant part of the time-dependence in the dissipative coefficients $\gamma(t)$ \cite{Shandera_2018}, including the periodicity. The background expansion may suggest the appropriate scale to use in an effective open model.

A further wrinkle in open effective theories is that the master equations need not be local in time. Most constructions in cosmology restrict to time-local equations for simlicity \cite{Agon2018,Shandera_2018,Brahma:2022yxu,Colas:2022hlq} although work in the context of black holes has shown that non-time-local evolution can be required \cite{Burgess:2021luo}. Our work here explores the conditions necessary for time-local dynamics in small qubit models. We find that even in the regions of parameter space with the largest degrees of symmetry, it can be necessary to use a time-non-local master equation. We make this determination based on the non-invertibility of the dynamical map, which depends both on the Hamiltonian and the initial environment state. Strong coupling is a necessary condition for non-invertibility, however so is a rather restricted set of initial environment states. Those states are defined largely by the dynamics of the free Hamiltonian for the environment. Diagonal initial environment states in the eigenbasis of $H_{\rm E;\rm free}$ are the most non-local, as their dynamical maps are non-invertible in the presence of strong coupling. 

In cosmology, the Hamiltonian is generically time-dependent, so it is not uncommon for the relevant (active) degrees of freedom and couplings to evolve over time. We used a 3-qubit model to investigate how such a time-dependence effects the nature of the dynamical map. We found that introducing a degree of freedom that does not respect the symmetry of the initial interaction drastically altered the reduced dynamics. The dynamical map became non-invertible at all times after the activation of the spectator qubit. This strongly suggests that the framework of open effective theories in cosmology should include a time-non-local component.

\acknowledgments
 We thank Nishant Agarwal, Sarang Gopalakrishnan, Archana Kamal, and Sahin Ozdemir for helpful discussions in the construction of this work. We further would like to acknowledge the useful suggestions made by Unnati Akhouri, Brenden Bowen, and Andrew Keefe. This work was supported by the National Science Foundation under PHY-1719991.

 \appendix

\bibliographystyle{unsrt}
\bibliography{dynmap}

\appendix

\section{Derivation of the maximally entangling parameter space}
\label{appendix:entangle}
 Following \cite{Makhlin:2000, Zhang:2003zz}, we find the essential non-local properties of $U$ by changing to the Bell basis using the unitary operator, 
\begin{equation}
    Q =\frac{1}{\sqrt{2}}\begin{bmatrix} 1&0&0&i \\0&i&1&0 \\0&i&-1&0 \\1&0&0&-i \end{bmatrix} \,.
\end{equation}

 The choice of Bell states is made so that the entanglement quadratic form $\hat{Ent}$ maps to the identity matrix. The operator $\hat{Ent}$ is defined as, 
 \begin{equation}
     Det I_{\rm C}(|\psi\rangle)=\psi_{00}\psi_{11}-\psi_{01}\psi_{10}= |\psi\rangle^{T}\hat{Ent} |\psi \rangle \, ,
 \end{equation}
 where $I_{\rm C}$ is the Choi isomorphism $I_{\rm C}: \mathcal{H} \rightarrow Mat(2,\mathbb{C})$, and one sees upon further inspection that $\hat{Ent}=-\frac{1}{2} Y_{\rm S}\otimes Y_{\rm E}$ and $Q^{T}(\hat{Ent})Q=\frac{1}{2}\mathbb{1}_{\rm S}\otimes\mathbb{1}_{\rm E}$. At the level of observables $Q$ takes the subalgebra of local observables $\mathfrak{su}(2)_{\rm S}\oplus \mathfrak{su}(2)_{\rm E}$ to the generators of 4D rotations on the Bell space $\mathfrak{so}(4)_{\rm Bell}$. 
 
 The time evolution operator is expressed in the Bell basis as, 
\begin{equation}
   U_{\rm Bell }(t)=\begin{bmatrix} \cos\omega_{+}t &0&0&e^{i\phi_{+}}\sin\omega_{+}t\\ 0&\cos\omega_{-}t&-e^{ i\phi_{-}}\sin\omega_{-}t&0\\0& e^{ -i\phi_{-}} \sin\omega_{-}t & \cos\omega_{-}t&0\\-e^{-i\phi_{+}}\sin\omega_{+}t&0&0&\cos\omega_{+}t\end{bmatrix}
\end{equation}
 The non-local properties of the time evolution in the reduction frame are determined by the eigenvalues of  $U_{\rm Bell}^{T}U_{\rm Bell}$ found to be, 
 \begin{equation}
 \begin{split}
     &u_{+}=\cos^{2}\omega_{+}t+(\cos2\phi_{+}+i\sqrt{\sin^{2}2\phi_{+}+\sin^{2}\phi_{+}})\sin^{2}\omega_{+}t \\ 
     &u_{-}=\cos^{2}\omega_{+}t+(\cos2\phi_{+}-i\sqrt{\sin^{2}2\phi_{+}+\sin^{2}\phi_{+}})\sin^{2}\omega_{+}t \\ 
     &v_{+}=\cos^{2}\omega_{-}t+(\cos2\phi_{-}+i\sqrt{\sin^{2}2\phi_{-}+\sin^{2}\phi_{-}})\sin^{2}\omega_{-}t \\ 
     &v_{-}= \cos^{2}\omega_{-}t+(\cos2\phi_{-}-i\sqrt{\sin^{2}2\phi_{-}+\sin^{2}\phi_{-}})\sin^{2}\omega_{-}t 
  \end{split}\,.
 \end{equation}

Assume we are given a linear combination of these eigenvalues $ a u_{+}+b u_{-}+c v_{+}+d v_{-}$ such that  $ a, b, c, d \geq 0$ and $a+b+c+d=1$. For this combination to be real we must have $a=b$ and $c=d$. Setting $a=\cos^{2}\varphi$ and $c=\sin^{2}\varphi$ we have, 
\begin{equation}
\begin{split}
   0&= \cos^{2}\varphi(\cos^{2}\omega_{+}t+\cos2\phi_{+}\sin^{2}\omega_{+}t)+\sin^{2}\varphi(\cos^{2}\omega_{-}t+\cos2\phi_{-}\sin^{2}\omega_{-}t) \\ &=\cos^{2}\varphi(|\alpha_{+}|^{2}(t)-\beta^{2}_{+}(t))+\sin^{2}\varphi(|\alpha_{-}|^{2}(t)-\beta^{2}_{-}(t))
\end{split}
\end{equation}
 A little bit of algebra and we find, 
 \begin{equation}
     \cos^{2}\varphi \beta_{+}^{2}(t)+\sin^{2}\varphi \beta_{-}^{2}(t)=\frac{1}{2}\,.
 \end{equation}
This condition may only be satisfied iff the largest of $\beta^{2}_{+}(t)$ and $\beta^{2}_{-}(t)$ is greater than or equal to $\frac{1}{2}$, or equivalently Max$(\phi_{+},\phi_{-})\geq \frac{\pi}{4}$. That is maximally entangled states may only be generated when $U$ has large enough off diagonal components, which we see becomes one of the conditions that non-invertibilities appear in the reduced dynamics.

\section{Divisibility in channels}
\label{appendix:divisibility}
 
In this appendix we establish the divisibility of the dynamical map family $\mathcal{D}_{+}$. For simplicity we assume that $\vec{r}_{E}(0)=\vec{0}$ although the results derived apply even if $\vec{r}_{\rm E}(0)=x_{\rm E}(0)\hat{x}$. We are interested in when the map $\Phi(\tau_{2},\tau_{1})$ fails to be completely positive, and how this depends on $\tau_{2}$ and $\phi$. To that end, the dynamical map has the structure, 
\begin{equation}
    \Lambda(t)=\begin{bmatrix} 1&0&0&0 \\0&\Lambda_{xx}(t)&\Lambda_{xy}(t)&0\\0&\Lambda_{yx}(t)&\Lambda_{yy}(t)&0\\0&0&0&\Lambda_{zz}(t)\end{bmatrix}
\end{equation}
with determinant Det$\Lambda=\Lambda^{2}_{zz}$ and inverse,
\begin{equation}
    \Lambda^{-1}(t)=\begin{bmatrix} 1&0&0&0 \\0&\frac{\Lambda_{yy}(t)}{\Lambda_{zz}(t)}&-\frac{\Lambda_{xy}(t)}{\Lambda_{zz}(t)}&0\\0&-\frac{\Lambda_{yx}(t)}{\Lambda_{zz}(t)}&\frac{\Lambda_{xx}(t)}{\Lambda_{zz}(t)}&0\\0&0&0&\frac{1}{\Lambda_{zz}(t)}\end{bmatrix} \,,
\end{equation}
and the (possibly singular) interweaving map is, 
\begin{equation}
    \Phi(\tau_{2},\tau_{1})=\begin{bmatrix} 1&0&0&0 \\0&\frac{\Lambda_{xx}(\tau_{2})\Lambda_{yy}(\tau_{1})-\Lambda_{xy}(\tau_{2})\Lambda_{yx}(\tau_{1})}{\Lambda_{zz}(\tau_{1})}&\frac{-\Lambda_{xx}(\tau_{2})\Lambda_{xy}(\tau_{1})+\Lambda_{xy}(\tau_{2})\Lambda_{xx}(\tau_{1})}{\Lambda_{zz}(\tau_{1})}&0\\0&\frac{\Lambda_{yx}(\tau_{2})\Lambda_{yy}(\tau_{1})-\Lambda_{yy}(\tau_{2})\Lambda_{yx}(\tau_{1})}{\Lambda_{zz}(\tau_{1})}&\frac{\Lambda_{yy}(\tau_{2})\Lambda_{xx}(\tau_{1})-\Lambda_{yx}(\tau_{2})\Lambda_{xy}(\tau_{1})}{\Lambda_{zz}(\tau_{1})}&0\\0&0&0&\frac{\Lambda_{zz}(\tau_{2})}{\Lambda_{zz}(\tau_{1})}\end{bmatrix} 
      \,. 
\end{equation}
N.B. that the map $\Phi(\tau_{2},\tau_{1})$ can have a restricted domain, where instead of the Bloch ball the interweaving map only acts on the image of $\Lambda(\tau_{1})$.

The criterion in \cite{Ruskai2001} asserts that $\Phi(\tau_{2},\tau_{1})$ is completely positive if, 
\begin{equation}
\begin{split}
    &\left[\frac{\Lambda_{xx}(\tau_{2})\Lambda_{yy}(\tau_{1})-\Lambda_{xy}(\tau_{2})\Lambda_{yx}(\tau_{1})}{\Lambda_{zz}(\tau_{1})}\pm\frac{\Lambda_{yy}(\tau_{2})\Lambda_{xx}(\tau_{1})-\Lambda_{yx}(\tau_{2})\Lambda_{xy}(\tau_{1})}{\Lambda_{zz}(\tau_{1})}\right]^{2} 
    \\\\ &\leq \left[1\pm\frac{\Lambda_{zz}(\tau_{2})}{\Lambda_{zz}(\tau_{1})}\right]^{2} \,.
\end{split}
\end{equation}
The + inequality is saturated at all times, so the components of $\Lambda$ satisfy the relation (recall $\Lambda_{yx}(t)=-\Lambda_{xy}(t)$),
\begin{equation}
    \Lambda_{xx}(\tau_{2})\Lambda_{yy}(\tau_{1})+2\Lambda_{xy}(\tau_{2})\Lambda_{xy}(\tau_{1})+\Lambda_{yy}(\tau_{2})\Lambda_{xx}(\tau_{1})=\Lambda_{zz}(\tau_{2})+\Lambda_{zz}(\tau_{1}) \,.
\end{equation}
Thus $\Phi$ is completely positive if, 
\begin{equation}
    (\Lambda_{xx}(\tau_{2})\Lambda_{yy}(\tau_{1})-\Lambda_{yy}(\tau_{2})\Lambda_{xx}(\tau_{1}))^{2} \leq (\Lambda_{zz}(\tau_{2})-\Lambda_{zz}(\tau_{1}))^{2}\,. 
\end{equation}
We already see that if $\tau_{n}=\frac{n\pi}{2\omega}$, then $\Phi(\tau_{n},\tau_{1})$ is CP as the above inequality reduces to a CP condition satisfied by $\Lambda(\tau_{1})$. Therefore, special times $\tau_{2}$ exist where the dynamics is CP divisible. This condition is not dependent on what particular values are chosen for $\omega$ and $\phi$. However for other values of $\tau_{2}$, the CP inequality will fail to be satisfied for certain values of $\tau_{1}$. The size of this interval is not dependent on $\omega$, but depends on $\phi$ and $\tau_{2}$. 
\end{document}